\documentclass[a4paper,12pt]{article}

\usepackage{amsmath}
\usepackage[psamsfonts]{amssymb}
\usepackage{rsfs}
\usepackage{bm}

\usepackage{cite}
\usepackage[dvips]{graphicx}
\usepackage{color}

\makeatletter
\@addtoreset{equation}{section}
\makeatother

\begin{document} 

\begin{titlepage}

\baselineskip 10pt
\hrule 
\vskip 5pt
\leftline{}
\leftline{Chiba Univ. Preprint
          \hfill   \small \hbox{\bf CHIBA-EP-141}}
\leftline{\hfill   \small \hbox{hep-th/0307270}}
\leftline{\hfill   \small \hbox{July 2003}}
\vskip 5pt
\baselineskip 14pt
%\leftline{}
\hrule 
\vskip 1.0cm
\centerline{\Large\bf 
Vacuum condensates, effective gluon mass
} 
\vskip 0.3cm
\centerline{\Large\bf  
and color confinement
}%$^*$
\vskip 0.3cm
\centerline{\Large\bf  
in a new reformulation of QCD
%in the large $N_c$ expansion
}%$^*$
\vskip 0.3cm
\centerline{\large\bf  
}%$^*$

\vskip 0.5cm

\centerline{{\bf 
Kei-Ichi Kondo$^{\dagger,{1}}$ 
}}  
\vskip 0.5cm
\centerline{\it
${}^{\dagger}$Department of Physics, Faculty of Science, 
Chiba University, Chiba 263-8522, Japan
%$\&$ Graduate School of Science and Technology, 
}
%\vskip 0.3cm
%\centerline{\it
%Chiba University, Chiba 263-8522, Japan
%}
\vskip 1cm
%\begin{description}
%\item[]{\it \centerline{  
%${}^{\dagger}$Department of Physics, Faculty of Science, 
%Chiba University,  Chiba 263-8522, Japan}
%}
%\vskip -0.5cm
%\item[]{\it 
%${}^{\ddagger}$Graduate School of Science and Technology, 
%Chiba University, Chiba 263-8522, Japan
%}
%\end{description}

\begin{abstract}
We propose a new reformulation of Yang-Mills theory in which three- and four-gluon self-interactions are eliminated at the price of introducing a sufficient number of auxiliary fields.  
We discuss the validity of this reformulation in the possible applications such as dynamical gluon mass generation, color confinement and glueball mass calculation. 
Moreover, we set up a new $1/N_c$ color expansion in the $SU(N_c)$ Yang-Mills theory based on this reformulation.  
In fact, we give the Feynman rules of the   $1/N_c$ expansion in the manifestly Lorentz covariant gauge. 
The Yang-Mills theory is defined on a non-trivial vacuum where  color--singlet transverse gluon pair condensations take place  by the attractive gluonic self-interactions.
This vacuum condensation provides a common non-vanishing mass for all the gluons with color symmetry being preserved.  
It is shown that the auxiliary fields become dynamical by acquiring the kinetic term due to quantum corrections. 
Then the static  potential between a pair of color charges is derived  as a combination of the Yukawa-type potential and the linear potential with non-vanishing string tension.
The mass of the lightest scalar glueball is calculated as the ratio to the gluon mass. 
The explicit calculations are performed as a partial resummation of the  leading order diagrams for the small 't~Hooft coupling. 

\end{abstract}

%\vskip 0.3cm
Key words:  vacuum condensation, large N expansion, mass gap, confinement, Yang--Mills theory, auxiliary field method, 

PACS: 12.38.Aw, 12.38.Lg 
%\vskip 0.2cm
\hrule  
\vskip 0.1cm
${}^1$ 
  E-mail:  {\tt kondok@faculty.chiba-u.jp}

\par 
\par\noindent
%\vskip 0.5cm

%%%%%%%%%%%%%%%%%%%%%%%%%%%%%%
%\vskip 2cm  
%\hrule  
%\bigskip  
%\centerline
%{\bf CHIBA UNIVERSITY}  
%\vfill 

%\hrule  
\vskip 0.5cm
%\begin{description}
%\item[]{
%$^\ddagger$
%  On leave of absence from: \\
%  Department of Physics, Faculty of Science,
%  Chiba University, Chiba 263, Japan.
%  }
%\item[]{
%$^*$ To be published in .
% Submitted to .
% }  
%\end{description}

\newpage
%\hrule  
%%%%% Table of Contents %%%%%
\pagenumbering{roman}
\tableofcontents
%%%%% Table of Contents %%%%%
%%%%%
%\baselineskip 23pt

\vskip 0.5cm  
%\hrule  

%%%%%%%%%%%%%%%%%%%%%%%%%%

%%%%%%%%%%%%%%%%%%%%%%%%%%

\end{titlepage}

%\newpage

\pagenumbering{arabic}

\baselineskip 14pt
%%%%%%%%%%%%%%%%%%%%%%%%%%%%%%%%%%%%%%%%%%%%%%%%%%%%%%%%%%%%%%%%%%%%%
%%%%%%%%%%%%%%%%%%%%%%%%%%%%%%%%%%%%%%%%%%%%%%%%%%%%%%%%%%%%%%%%%%%%%
%%%%%%%%%%%%%%%%%%%%%%%%%%%%%%%%%%%%%%%%%%%%%%%%%%%%%%%%%%%%%%%%%%%%%
%%%%%%%%%%%%%%%%%%%%%%%%%%%%%%%%%%%%%%%%%%%%%%%%%%%%%%%%%%%%%%%%%%%%%
%%%%%%%%%%%%%%%%%%%%%%%%%%%%%%%%%%%%%%%%%%%%%%%%%%%%%%%%%%%%%%%%%%%%%
%\newpage
\section{Introduction}

The mechanism for providing the gauge boson with the mass in strong interactions is still under the debate, 
although fifty years are going to pass since the discovery of the non-Abelian gauge theory by Yang and Mills \cite{YM54}.
In this paper we argue that 
the normal massless vacuum of Yang--Mills theory is unstable, since a group singlet pair of transverse Yang--Mills gauge boson forms a bound state with a negative energy eigenvalue (Cooper pair) due to attractive force. 
Consequently, the stable vacuum is realized after the condensation of these pairs.  
The excitation spectrum has a gap without breaking the global gauge (color) symmetry for any coupling strength and the gauge bosons acquire a common mass.  

In the classical level, Yang-Mills theory is a scale-invariant (actually, conformal invariant) field theory describing massless gauge bosons.  By the gauge principle, the existence of the mass term is prohibited in the classical Yang-Mills theory. 
In the quantum level, on the other hand, one has the possibility of the dynamical mass generation for the gauge boson as investigated long ago  \cite{Smit74,EF74,TTP75,FK78,Fukuda77,Fukuda78,GM78}.

In the above context, among others, it is important to recall a pioneering work by 
Fukuda and Kugo \cite{FK78} in which they discussed the transverse gluon pair condensation by using the variational technique for $SU(N)$ Yang--Mills theory already in the late 1970s. 
They adopted the Coulomb gauge in the Hamiltonian formalism and dealt  with only the transverse modes with no ghost. 
The basic tools were the Cooper equation, Bogoliubov transformation and the gap equation. 
However, the resulting energy spectrum $\Delta(p)$ did not have the relativistically invariant form $\sqrt{p^2+m^2}$, since their analyses are based on the non-covariant formalism. 
In the second paper \cite{Fukuda77}, Fukuda attacked the same problem in covariant form by making use of the effective action or potential and the resulting Schwinger-Dyson (SD) equation.  
He obtained qualitatively the same result as the first paper. 
These works demonstrated that the vacuum of the massive phase can have the lower energy than the normal vacuum. 

%%%%%%%%%%%%%%%%%%%%%%%%%%%%%%%%%%%%%%%%%%%%%%%%%%%%%%%%%%%%

\begin{figure}[htbp]
\begin{center}
\includegraphics[height=5cm]{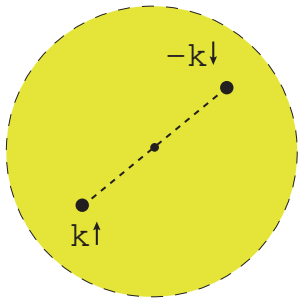}
\caption{Gluon pair condensation in color singlet and spin singlet channel (center of mass frame).
}
\label{fig:Pair_condensation}
\end{center}
\end{figure}

%%%%%%%%%%%%%%%%%%%%%%%%%%%%%%%%%%%%%%%%%%%%%%%%%%%%%%%%%%%%

If two gauge particles form a bound state, the bound state should be a tachyon pole, since two gauge particles are massless. 
The existence of the tachyon bound state was also discussed by solving the Bethe-Salpeter (BS) equation in the ladder approximation \cite{Fukuda78,GM78} where the Landau gauge $\lambda=0$ is a preferred choice. 
From the viewpoint of the dynamical symmetry breaking, the existence of Goldstone pole as a consequence of the condensation phenomena was discussed by several works \cite{Smit74,EF74,TTP75} in the framework of SD and BS equations.

However, it should be remarked that the solutions of the SD and BS equations strongly depend  on the choice of gauge fixing parameter $\lambda$.
In fact, the result is strongly gauge dependent; for $\lambda>-1$ the tachyon pole which is group singlet and spin singlet exists, whereas such a pole does not exist for $\lambda<-1$.
It is notorious that the systematic improvement is rather difficult in the SD and BS approach.

In this paper, we discuss the problem
based on a manifestly Lorentz covariant formulation.  
Therefore, we can obtain the manifestly relativistic result. Quite recently, it was shown \cite{Kondo03b} that the transverse gluon pair condensation can be replaced by the on-shell BRST invariant vacuum condensate of mass dimension 2 which was proposed by the author \cite{Kondo01,KMSI02}.  
In this paper, we show that the $1/N_c$ expansion can be set up to yield the condensation phenomena in the systematic way.  The $1/N_c$ expansion is a non-perturbative method which is systematically improvable, in contrast to solving the approximate gap equation by adopting the ad hoc approximation. 
In our approach, the existence of the pair condensation is also a gauge independent phenomena in the sense; if it occurs at a certain gauge at all, it should occur in all the other gauges.  

The $1/N$ expansion was extensively used in the past works. 
The well-known examples are among others the O(N) symmetric four-fermion model (e.g., two-dimensional Gross-Neveu model) and the  
non-linear $\sigma$ model \cite{DLV78}.  
It is also remarkable that the three-dimensional four-fermion model become renormalizable in the $1/N_f$ expansion \cite{RWP89}, although it is non-renormalizable in weak-coupling perturbation expansion.
The  four-dimensional Yang-Mills theory in the $1/N_c$ expansion might be related to the string theory \cite{tHooft74}.  
See e.g., \cite{largeN} for reviews of large N expansion.

It should be noted that the $1/N_c$ expansion given in this paper is quite different from the conventional one for the Yang-Mills theory \cite{tHooft74,largeN}.  In this paper we adopt the Lorentz covariant gauge.  The Maximal Abelian (MA) gauge will be treated in a separate paper \cite{Kondo03c}.

This paper is organized as follows. 
In section 2, we rewrite the original Yang-Mills action which includes the cubic and quartic gluon self-interactions to an equivalent theory without gluon self-interaction by introducing sufficient number of auxiliary fields.   Consequently, the gluon field can be integrated out exactly.  
In section 3, we set up a new $1/N_c$ expansion of the $SU(N_c)$ Yang-Mills theory on a non-trivial vacuum where the vacuum condensation of mass dimension two takes place. We give Feynman rules of the relevant $1/N_c$ expansion.  The low-energy and high-energy behavior of the auxiliary field propagators are studied.  
As a by-product, the mass of the lightest glueball is given. 
In section 4, we obtain the effective potential of the theory and estimate the effective gluon mass by making use of the lattice data.
In section 5, we calculate the static potential between a pair of color charges to the leading order of the $1/N_c$ expansion.  We show that the static potential is composed of the Yukawa-type potential and the linear potential with the non-vanishing string tension which is proportional to the vacuum condensate of mass dimension 2. 
In section 6, we discuss the string representation of the Yang-Mills theory within the scheme of the  $1/N_c$ expansion.  
The final section is devoted to conclusion and discussion.
Some of the technical materials are given in the Appendices A and B.

%%%%%%%%%%%%%%%%%%%%%%%%%%%%%%%%%%%%%%%%%%%%%%%%%%%%%%%%%%%%%%%%%%%%%
%%%%%%%%%%%%%%%%%%%%%%%%%%%%%%%%%%%%%%%%%%%%%%%%%%%%%%%%%%%%%%%%%%%%%
%%%%%%%%%%%%%%%%%%%%%%%%%%%%%%%%%%%%%%%%%%%%%%%%%%%%%%%%%%%%%%%%%%%%%
%%%%%%%%%%%%%%%%%%%%%%%%%%%%%%%%%%%%%%%%%%%%%%%%%%%%%%%%%%%%%%%%%%%%%
%%%%%%%%%%%%%%%%%%%%%%%%%%%%%%%%%%%%%%%%%%%%%%%%%%%%%%%%%%%%%%%%%%%%%

%\newpage
\section{Reformulation based on the auxiliary field method}
\subsection{Yang-Mills action}

The Euclidean version of the Yang-Mills Lagrangian is given by
\begin{align}
\mathscr{L}_{\text{YM}} 
 =&   \frac14 \left({\cal F}_{\mu\nu} \right)^2 
 =  {1 \over 4} (\partial_\mu \mathscr{A}_\nu - \partial_\nu \mathscr{A}_\mu + g \mathscr{A}_\mu \times \mathscr{A}_\nu)^2 ,
\nonumber\\
  =&  
+  {1 \over 4} (\partial_\mu \mathscr{A}_\nu - \partial_\nu \mathscr{A}_\mu)^2 
\nonumber\\
 &  +{1 \over 2} g (\partial_\mu \mathscr{A}_\nu - \partial_\nu \mathscr{A}_\mu) \cdot (\mathscr{A}^\mu \times \mathscr{A}^\nu) 
%\leftarrow \text{3-point interaction}
\nonumber\\
 &+{1 \over 4} g^2 (\mathscr{A}_\mu \times \mathscr{A}_\nu) \cdot (\mathscr{A}^\mu \times \mathscr{A}^\nu) ,
%\leftarrow \text{4-point interaction}
  \label{YM1}
\end{align}
where 
${\cal F}_{\mu\nu}^A$ is the non-Abelian field strength defined by
$
 {\cal F}_{\mu\nu}^A
  =\partial_\mu {\cal A}_\nu^A
   -\partial_\nu {\cal A}_\mu^A
   +gf^{ABC}{\cal A}_\mu^B{\cal A}_\nu^C  
$
with $A,B,C=1,\cdots,{\rm dim}(G)$ for the gauge group $G$. 
In what follows, we use a notation
$F^2 := F \cdot F $,  
$(F \cdot G) :=   F^A G^A = 2{\rm tr}(FG)$,  
$
  (F \times G)^A := f^{ABC} F^B G^C =-2i {\rm tr}(T^A [F,G])
$ 
with
the normalization for the generators $T^A$ of the Lie algebra $\mathscr{G}$ of the gauge group $G$: ${\rm tr}(T^A T^B)={1 \over 2}\delta^{AB}$.
Then the generating functional is given by the functional integral:
\begin{align}
  Z[J] := \int \mathcal{D}\mathscr{A}_\mu^A \mathcal{D}\mathscr{B}^A \mathcal{D}\mathscr{C}^A \mathcal{D}\bar{\mathscr{C}}^A \exp \left\{ -\int d^4x [\mathscr{L}_{\rm YM}^{\rm tot} + \mathscr{L}^{J}]\right\} ,
\label{Z[J]}
\end{align}
where we have included the source term 
$
\mathscr{L}^{J}=\mathscr{A}_\mu \cdot J^\mu + \mathscr{B} \cdot J_B + J_c \cdot \mathscr{C} + J_{\bar{c}} \cdot \bar{\mathscr{C}}
$.
Here the total Lagrangian of the Yang-Mills theory 
$
  \mathscr{L}_{\rm YM}^{\rm tot} = \mathscr{L}_{\rm YM}   + \mathscr{L}_{\rm GF+FP} ,
$
is given by adding the gauge fixing (GF) and Faddeev-Popov (FP) ghost term $\mathscr{L}_{\rm GF+FP}$ to the pure Yang-Mills term $\mathscr{L}_{\rm YM}$ where $\mathscr{B}^A$ is the Nakanishi-Lautrup field, and $\mathscr{C}^A$ and $\bar{\mathscr{C}}^A$ are the FP ghost and antighost fields.  

%\marginpar{1x1=2+1+0}

First, we introduce the second-rank antisymmetric tensor field $B_{\mu\nu}$ as the Hodge dual of the distorted field strength according to \cite{QR97,KondoI,KS01}%
%\footnote{
%The unequal treatment of $(\partial_\mu \mathscr{A}_\nu - \partial_\nu \mathscr{A}_\mu)$ and $g(\mathscr{A}_\mu \times \mathscr{A}_\nu)$ breaks the local gauge invariance.  However, it should be noted that the local gauge invariance is sooner or later broken by the gauge fixing.  On the other hand, the global gauge invariance is maintained even after this procedure. 
%}
\begin{align}
    B_{\mu\nu} \leftrightarrow {}^*[\rho (\partial_\mu \mathscr{A}_\nu - \partial_\nu \mathscr{A}_\mu) + \sigma g (\mathscr{A}_\mu \times \mathscr{A}_\nu)] ,
\end{align}
where $\rho$ and $\sigma$ are two parameters to be determined later 
and the Hodge star (duality) operation ${}^*$ for the second-rank tensor is defined by
$
  {}^*H_{\mu\nu} := \frac{1}{2} \epsilon_{\mu\nu\rho\sigma} H^{\rho\sigma}  
$.
For this purpose, we insert an  identity into the functional integral (\ref{Z[J]}):
\begin{align}
  1 = \int \mathcal{D}B_{\mu\nu}^A \exp \left[ - \int d^4x 
\frac{1}{4} \left\{ B_{\mu\nu}^A-i {}^*[\rho (\partial_\mu \mathscr{A}_\nu - \partial_\nu \mathscr{A}_\mu)^A+ \sigma g (\mathscr{A}_\mu \times \mathscr{A}_\nu)^A] \right\}^2 
\right] .
\label{id-B}
\end{align}
Then the Lagrangian is modified as 
\begin{align}
  \mathscr{L}_{\text{YM}} 
  =&  
\frac{1}{4} B_{\mu\nu} \cdot B^{\mu\nu} - {i \over 2} B^{\mu\nu} \cdot {}^*[\rho (\partial_\mu \mathscr{A}_\nu - \partial_\nu \mathscr{A}_\mu) + \sigma g (\mathscr{A}_\mu \times \mathscr{A}_\nu)] 
\nonumber\\ &
+ {1-\rho^2 \over 4} (\partial_\mu \mathscr{A}_\nu - \partial_\nu \mathscr{A}_\mu)^2 
\nonumber\\
 &  +{1-\rho \sigma \over 2} g (\partial_\mu \mathscr{A}_\nu - \partial_\nu \mathscr{A}_\mu) \cdot (\mathscr{A}^\mu \times \mathscr{A}^\nu) 
\nonumber\\
 &+{1-\sigma^2 \over 4} g^2 (\mathscr{A}_\mu \times \mathscr{A}_\nu) \cdot (\mathscr{A}^\mu \times \mathscr{A}^\nu) .
 \label{YM2}
\end{align}
For any values  of the two parameters $\rho$ and $\sigma$, the Lagrangian (\ref{YM2}) is equivalent to the original Lagrangian (\ref{YM1}). 
Note that $B_{\mu\nu}$ is the auxiliary field without its kinetic term.

Next, we rewrite the last term in (\ref{YM2})  which is quartic in the gluon field $\mathscr{A}_\mu$.
For $SU(N_c)$, the quartic term has another expression%
\footnote{
We have used an identity for the structure constant (see \cite{MSW68} for a derivation):
\begin{equation}
 f^{ABE}f^{CDE} = {2 \over N_c}(\delta^{AC}\delta^{BD}-\delta^{AD}\delta^{BC}) +  d^{ACE}d^{BDE} - d^{BCE}d^{ADE}  . 
\end{equation}
For the generator $T^A$ of the Lie algebra $\mathscr{G}$ of the gauge group $G$, 
$[T^A, T^B] = if^{ABC} T^C$ and
$\{ T^A, T^B \} = {1 \over N} \delta^{AB} + d^{ABC} T^C$.
}
\begin{align}
  (\mathscr{A}_\mu \times \mathscr{A}_\nu) \cdot (\mathscr{A}^\mu \times \mathscr{A}^\nu)
  =& {2 \over N_c} \left[ (\mathscr{A}_\mu \cdot \mathscr{A}^\mu)(\mathscr{A}_\nu \cdot \mathscr{A}^\nu) - (\mathscr{A}_\mu \cdot \mathscr{A}_\nu)(\mathscr{A}^\mu \cdot \mathscr{A}^\nu) \right] 
  \nonumber\\&
  + (\mathscr{A}_\mu \star \mathscr{A}^\mu) \cdot (\mathscr{A}_\nu \star \mathscr{A}^\nu) - (\mathscr{A}_\mu \star \mathscr{A}_\nu) \cdot (\mathscr{A}^\mu \star \mathscr{A}^\nu) ,
  \label{id}
\end{align}
where we have defined a new product 
$
  (F \star G)^A :=  d^{ABC} F^B G^C  
= 2 {\rm tr}(T^A \{ F, G \})
$
with completely symmetric tensor $d^{ABC}$ for $N_c \ge 3$.   
For $SU(2)$, we do not need to consider the last two terms in (\ref{id}), since $d^{ABC}=0$ for $SU(2)$, and hence the quartic term is greatly simplified \cite{DK03}:
\begin{align}
  (\mathscr{A}_\mu \times \mathscr{A}_\nu) \cdot (\mathscr{A}^\mu \times \mathscr{A}^\nu)
  =&  (\mathscr{A}_\mu \cdot \mathscr{A}^\mu)(\mathscr{A}_\nu \cdot \mathscr{A}^\nu) - (\mathscr{A}_\mu \cdot \mathscr{A}_\nu)(\mathscr{A}^\mu \cdot \mathscr{A}^\nu) .
  \label{id2}
\end{align}

For later convenience, the identity (\ref{id}) is rewritten into
\begin{align}
  (\mathscr{A}_\mu \times \mathscr{A}_\nu) \cdot (\mathscr{A}^\mu \times \mathscr{A}^\nu)
  =&  {3 \over 2N_c}(\mathscr{A}_\mu \cdot \mathscr{A}^\mu)(\mathscr{A}_\nu \cdot \mathscr{A}^\nu) - {2 \over N_c} S_{\mu\nu} S^{\mu\nu} 
  \nonumber\\&
  + {3 \over 4}(\mathscr{A}_\mu \star \mathscr{A}^\mu)^2  - T_{\mu\nu} \cdot T^{\mu\nu}  ,
  \label{id3}
\end{align}
by introducing the {\it traceless} symmetric tensors $S_{\mu\nu}=S_{\nu\mu}$  and $T^A_{\mu\nu}=T^A_{\nu\mu}$ defined by 
\begin{align}
  S_{\mu\nu} :=& (\mathscr{A}_\mu \cdot \mathscr{A}_\nu) - {1 \over 4} \delta_{\mu\nu} (\mathscr{A}_\rho \cdot \mathscr{A}^\rho), 
  \quad S_\mu{}^\mu = 0 ,
\\
  T^A_{\mu\nu} :=& (\mathscr{A}_\mu \star \mathscr{A}_\nu)^A - {1 \over 4} \delta_{\mu\nu} (\mathscr{A}_\rho \star \mathscr{A}^\rho)^A , 
  \quad T^A{}_\mu{}^\mu = 0 .
\end{align}

Moreover, we introduce the auxiliary scalar field $\phi$ and the auxiliary symmetric tensor fields $G_{\mu\nu}$ for SU(2) and also $\varphi^A$ and $V^A_{\mu\nu}$ for $SU(N_c)$ ($N_c \ge 3$). 
For example, the auxiliary field $\phi$ is introduced by inserting one more identity into (\ref{Z[J]})
\begin{subequations}
\begin{align}
  1 = \int \mathcal{D}\phi \exp \left\{ - \int d^4x 
{\sigma_\phi \over 2} \left( \phi + {1 \over 2} \mathscr{A}_\mu \cdot \mathscr{A}^\mu \right)^2 
\right\} ,
\label{aux-phi}
\end{align}
where $\sigma_\phi$ is an arbitrary parameter at this stage. 
Other auxiliary fields $G_{\mu\nu},\varphi^A$ and $V^A_{\mu\nu}$ are introduced in the similar way. 
\begin{align}
  1 =& \int \mathcal{D}G_{\mu\nu} \exp \left\{ - \int d^4x 
{\sigma_G \over 2} \left( G_{\mu\nu} + {1 \over 2} S_{\mu\nu} \right)^2 
\right\} ,
\\
  1 =& \int \mathcal{D}\varphi^A \exp \left\{ - \int d^4x 
{\sigma_\varphi \over 2} \left( \varphi^A + {1 \over 2} (\mathscr{A}_\mu \star \mathscr{A}^\mu)^A \right)^2 
\right\} ,
\\
  1 =& \int \mathcal{D}V_{\mu\nu}^A \exp \left\{ - \int d^4x 
{\sigma_V \over 2} \left( V_{\mu\nu}^A + {1 \over 2} T_{\mu\nu}^A \right)^2 
\right\} .
\end{align}
\end{subequations}
Then the quartic interaction in (\ref{YM2}) is replaced as follows. 
\begin{align}
  & {1-\sigma^2 \over 4} g^2 (\mathscr{A}_\mu \times \mathscr{A}_\nu) \cdot (\mathscr{A}^\mu \times \mathscr{A}^\nu) 
  \nonumber\\ &
  \rightarrow 
    {\sigma_\phi \over 2} \phi^2 + {\sigma_\phi \over 2} \phi (\mathscr{A}_\mu \cdot \mathscr{A}^\mu)
+ {\sigma_\phi+3(1-\sigma^2)g^2/N_c \over 8}(\mathscr{A}_\mu \cdot \mathscr{A}^\mu)^2
  \nonumber\\ &
  + {\sigma_G \over 2} G_{\mu\nu} G^{\mu\nu} + {\sigma_G \over 2} G_{\mu\nu} S^{\mu\nu}
+ {\sigma_G-4(1-\sigma^2)g^2/N_c \over 8}S_{\mu\nu} S^{\mu\nu}
  \nonumber\\ &
  + {\sigma_\varphi \over 2} \varphi \cdot \varphi + {\sigma_\varphi \over 2} \varphi \cdot (\mathscr{A}_\mu \star \mathscr{A}^\mu)
+ {\sigma_\varphi+(3/2)(1-\sigma^2)g^2 \over 8} (\mathscr{A}_\mu \star \mathscr{A}^\mu)^2
  \nonumber\\ &
  + {\sigma_V \over 2} V_{\mu\nu} \cdot V^{\mu\nu} + {\sigma_V \over 2} V_{\mu\nu} \cdot T^{\mu\nu}
+ {\sigma_V-2(1-\sigma^2)g^2 \over 8}T_{\mu\nu} \cdot T^{\mu\nu}  .
\label{Aux}
\end{align}

\subsection{Lorentz gauge fixing and FP ghost}

We adopt  the generalized Lorentz gauge or Curci-Ferrari (CF) gauge  \cite{CF76,BT82,Baulieu85,DJ82}.
The gauge fixing (GF) and the Faddeev-Popov (FP) ghost term is given by 
\begin{align}
  \mathscr{L}_{\rm GF+FP} &= \bm{\delta}_{\rm B} \bar{\bm{\delta}}_{\rm B} \left(  {1 \over 2} \mathscr{A}_\mu \cdot \mathscr{A}_\mu  + {\alpha' \over 2} \mathscr{C} \cdot \bar{\mathscr{C}} \right) 
- {\alpha \over 2} \mathscr{B} \cdot \mathscr{B} 
\\
&=  \bm{\delta}_{\rm B} \left( \partial_\mu \bar{\mathscr{C}} \cdot \mathscr{A}_\mu + {\alpha' \over 2} \bar{\mathscr{C}} \cdot \mathscr{B} - {1 \over 4} \alpha' g \bar{\mathscr{C}} \cdot (\bar{\mathscr{C}} \times \mathscr{C})   \right) 
- {\alpha \over 2} \mathscr{B} \cdot \mathscr{B}  .
\label{GFglobal2}
\end{align}
The conventional Lorentz covariant gauge corresponds to  $\alpha'=0$.
Performing the BRST  transformation (Euclidean version), 
\begin{subequations}
\begin{align}
 \bm{\delta}_{\rm B} \mathscr{A}_\mu(x)
   & =\mathscr{D}_\mu[\mathscr{A}]\mathscr{C}(x)
    :=\partial_\mu \mathscr{C}(x)
      + g (\mathscr{A}_\mu(x) \times \mathscr{C}(x)) , \\
 \bm{\delta}_{\rm B} \mathscr{C}(x)
   & =-{1 \over 2}g(\mathscr{C}(x) \times \mathscr{C}(x)) , \\
 \bm{\delta}_{\rm B} \bar{\mathscr{C}}(x)
   & = - \mathscr{B}(x) , \\
 \bm{\delta}_{\rm B} \mathscr{B}(x)
   &=0 ,
\label{BRST13}
\end{align}
\end{subequations}
and  anti-BRST transformation (Euclidean version), 
\begin{subequations}
\begin{align}
 \bar{\bm{\delta}}_{\rm B} \mathscr{A}_\mu(x) &=
\mathscr{D}_\mu[\mathscr{A}] \bar{\mathscr{C}}(x) :=
\partial_\mu \bar{\mathscr{C}}(x) + g (\mathscr{A}_\mu(x) \times \bar{\mathscr{C}}(x)) , \\
 \bar{\bm{\delta}}_{\rm B} \bar{\mathscr{C}}(x) &=-{1 \over 2}g(\bar{\mathscr{C}}(x) \times \bar{\mathscr{C}}(x)) ,
 \\
 \bar{\bm{\delta}}_{\rm B} \mathscr{C}(x) &= - \bar{\mathscr{B}}(x) ,
\\
 \bar{\bm{\delta}}_{\rm B} \bar{\mathscr{B}}(x) &=0 ,
\label{BRST23}
\end{align}
\end{subequations}
where
$
  \bar{\mathscr{B}}(x) =-\mathscr{B}(x) +  g (\mathscr{C}(x) \times \bar{\mathscr{C}}(x))  
$,
we obtain
\begin{align}
 \mathscr{L}_{\rm GF+FP} =\,& 
 - {\alpha+\alpha' \over 2} \mathscr{B} \cdot \mathscr{B} 
+ \mathscr{B} \cdot \partial_\mu \mathscr{A}_\mu
+ {\alpha' \over 2}  g (\mathscr{C} \times \bar{\mathscr{C}}) \cdot \mathscr{B} 
\nonumber\\&
 +  \bar{\mathscr{C}} \cdot \partial_\mu \mathscr{D}_\mu[\mathscr{A}]\mathscr{C}
+ {\alpha' \over 8}g^2 (\bar{\mathscr{C}} \times \bar{\mathscr{C}}) \cdot (\mathscr{C} \times \mathscr{C}) 
\\
=\,& - {\alpha+\alpha' \over 2}\mathscr{B} \cdot \mathscr{B}
+ \mathscr{B} \cdot \partial_\mu \mathscr{A}_\mu 
   + {\alpha' \over 2}  g (\mathscr{C} \times \bar{\mathscr{C}}) \cdot \mathscr{B} 
\nonumber\\&
+  \bar{\mathscr{C}} \cdot \partial_\mu \mathscr{D}_\mu[\mathscr{A}]\mathscr{C}
- {\alpha' \over 4}g^2 ( \mathscr{C} \times \bar{\mathscr{C}}) \cdot ( \mathscr{C} \times \bar{\mathscr{C}}) .
\end{align}
 The GF+FP term is further rewritten as
\begin{subequations}
\begin{align}
 S_{\rm GF+FP} 
=&  \int d^4x \Big\{ 
  {1 \over 2\lambda}(\partial_\mu \mathscr{A}_\mu)^2
+ (1-\xi) \bar{\mathscr{C}} \cdot \partial_\mu \mathscr{D}_\mu[\mathscr{A}]\mathscr{C}
+ \xi  \bar{\mathscr{C}} \cdot \mathscr{D}_\mu[\mathscr{A}] \partial_\mu \mathscr{C}
\nonumber\\&
- {1 \over 2}\lambda \xi (1-\xi) g^2 ( \mathscr{C} \times \bar{\mathscr{C}}) \cdot ( \mathscr{C} \times \bar{\mathscr{C}}) 
\nonumber\\&
 - {\lambda \over 2} \left( \mathscr{B} - \lambda^{-1} \partial_\mu \mathscr{A}_\mu - \xi g (\mathscr{C} \times \bar{\mathscr{C}}) \right)^2 \Big\} ,
\end{align}
or
\begin{align}
 S_{\rm GF+FP} 
=& \int d^4x \Big\{ {1 \over 2\lambda}(\partial_\mu \mathscr{A}_\mu)^2
+  \bar{\mathscr{C}} \cdot \partial_\mu \partial_\mu \mathscr{C}
+ (1-\xi)g    \mathscr{A}_\mu \cdot (\partial_\mu \bar{\mathscr{C}} \times \mathscr{C})
\nonumber\\&
- \xi g  \mathscr{A}_\mu \cdot (\bar{\mathscr{C}} \times \partial_\mu \mathscr{C})
- {1 \over 2}\lambda \xi (1-\xi) g^2 ( \mathscr{C} \times \bar{\mathscr{C}}) \cdot ( \mathscr{C} \times \bar{\mathscr{C}})
\nonumber\\&
 - {\lambda \over 2} \left( \mathscr{B} - \lambda^{-1} \partial_\mu \mathscr{A}_\mu - \xi g (\mathscr{C} \times \bar{\mathscr{C}}) \right)^2 \Big\} ,
 \label{GF+FP2}
\end{align}
where we have defined the two parameters 
\begin{equation}
  \lambda :=\alpha+\alpha' , \quad
  \xi :={\alpha'/2 \over \alpha+\alpha'} ={\alpha' \over 2\lambda}.
\end{equation}
\end{subequations}
In this form, it is easy to eliminate the Nakanishi-Lautrup field $\mathscr{B}$. 

For the Landau gauge $\lambda=0$, the ghost self-interaction disappears.  In the gauges other than the Landau $\lambda\not=0$, we  treat the four ghost interaction in the similar way to the four gluon interaction:  
\begin{align}
  (\bar{\mathscr{C}} \times \mathscr{C}) \cdot (\bar{\mathscr{C}} \times \mathscr{C})
= {2 \over N_c} (\bar{\mathscr{C}} \cdot \mathscr{C})^2 
+ (\bar{\mathscr{C}} \star \mathscr{C}) \cdot (\bar{\mathscr{C}} \star \mathscr{C}) ,
\end{align}
where we have used a fact that 
$\mathscr{C} \cdot \mathscr{C}=0=\bar{\mathscr{C}} \cdot \bar{\mathscr{C}}$ and
$\mathscr{C} \star \mathscr{C}=0=\bar{\mathscr{C}} \star \bar{\mathscr{C}}$ 
due to Grassmannian property. 
By introducing the auxiliary fields $\phi_{FP}$ and $\varphi_{FP}^A$ through the identities:
\begin{subequations}
\begin{align}
  1 =& \int \mathcal{D}\phi_{FP} \exp \left\{ - \int d^4x 
{\sigma_{FP} \over 2} \left( \phi_{FP} + \bar{\mathscr{C}} \cdot \mathscr{C}  \right)^2 
\right\} ,
\\
  1 =& \int \mathcal{D}\varphi_{FP}^A \exp \left\{ - \int d^4x 
 {\sigma_{FP}' \over 2} \left( \varphi_{FP} + \bar{\mathscr{C}} \star \mathscr{C}  \right)^2 \right\} ,
\end{align}
\end{subequations}
the four ghost interaction term is cast into the form:
\begin{align}
 & -{1 \over 2}\lambda \xi (1-\xi) g^2 ( \bar{\mathscr{C}} \times \mathscr{C}) \cdot ( \bar{\mathscr{C}} \times \mathscr{C})
+ {\sigma_{FP} \over 2} \left( \phi_{FP} + \bar{\mathscr{C}} \cdot \mathscr{C}  \right)^2 
+ {\sigma_{FP}' \over 2} \left( \varphi_{FP} + \bar{\mathscr{C}} \star \mathscr{C}  \right)^2 
\nonumber\\
=&  {\sigma_{FP} \over 2} \phi_{FP}^2 
+ {\sigma_{FP}' \over 2} \varphi_{FP} \cdot \varphi_{FP}   
+ \sigma_{FP} \phi_{FP}  \bar{\mathscr{C}} \cdot \mathscr{C} 
 + \sigma_{FP}' \varphi_{FP} \cdot (\bar{\mathscr{C}} \star \mathscr{C})
\nonumber\\&
+{1 \over 2} \left[ \sigma_{FP} - \lambda \xi (1-\xi) g^2{2 \over N_c} \right] (\bar{\mathscr{C}} \cdot \mathscr{C})^2 
+ {1 \over 2} \left[ \sigma_{FP}' - \lambda \xi (1-\xi) g^2 \right] (\bar{\mathscr{C}} \star \mathscr{C}) \cdot (\bar{\mathscr{C}} \star \mathscr{C}) .
\end{align}

 The auxiliary fields must be introduced such that the BRST symmetry is maintained. 
For this purpose, the simplest choice of the BRST transformation for the auxiliary fields is given as 
\begin{subequations}
\begin{align}
   \bm{\delta}_{\rm B} \phi =& -  \bm{\delta}_{\rm B}  \left( {1 \over 2} \mathscr{A}_\mu \cdot \mathscr{A}^\mu \right) 
%= - \mathscr{A}_\mu \cdot \partial^\mu \mathscr{C} 
%= \partial^\mu \mathscr{A}_\mu \cdot  \mathscr{C}  
%- \partial^\mu( \mathscr{A}_\mu \cdot \mathscr{C}) ,
\\
   \bm{\delta}_{\rm B} B_{\mu\nu} =& -  \bm{\delta}_{\rm B}  \left( {}^*[\rho (\partial_\mu \mathscr{A}_\nu - \partial_\nu \mathscr{A}_\mu) + \sigma g (\mathscr{A}_\mu \times \mathscr{A}_\nu)] \right) ,
\\
   \bm{\delta}_{\rm B} G_{\mu\nu} =& - \bm{\delta}_{\rm B}  \left( {1 \over 2} S_{\mu\nu} \right) ,
\\
   \bm{\delta}_{\rm B} \varphi =& - \bm{\delta}_{\rm B}  \left( {1 \over 2} \mathscr{A}_\mu \star \mathscr{A}^\mu \right) ,
\\
   \bm{\delta}_{\rm B} V_{\mu\nu} =& - \bm{\delta}_{\rm B}  \left(  {1 \over 2} T_{\mu\nu} \right) ,
\\
   \bm{\delta}_{\rm B} \phi_{FP} =& - \bm{\delta}_{\rm B} \left(  \bar{\mathscr{C}} \cdot \mathscr{C}  \right) ,
\\
   \bm{\delta}_{\rm B} \varphi_{FP} =& - \bm{\delta}_{\rm B} \left(  \bar{\mathscr{C}} \star \mathscr{C}  \right) .
\end{align}
\end{subequations}
The original measure 
$
 d\mu(\Phi) :=\mathcal{D}\mathscr{A}_{\mu} \mathcal{D}\mathscr{C} \mathcal{D}\bar{\mathscr{C}} \mathcal{D}\mathscr{B} 
$
is known to be BRST invariant  \cite{FS80}. 
Therefore the additional measure 
$
 d\mu(\Psi) := \mathcal{D}B_{\mu\nu} \mathcal{D}\phi \mathcal{D}G_{\mu\nu} \mathcal{D}\varphi \mathcal{D}V_{\mu\nu} \mathcal{D}\phi_{FP} \mathcal{D}\varphi_{FP} 
$
of the auxiliary fields is also BRST invariant, since 
it is easy to show that the Jacobian of the BRST transformation associated to the total measure $d\mu(\Phi)d\mu(\Psi)$ is equal to one. 
In particular, we find 
\begin{align}
   \bm{\delta}_{\rm B} \phi(x) 
= - \mathscr{A}_\mu(x) \cdot \partial^\mu \mathscr{C}(x)
= \partial^\mu \mathscr{A}_\mu(x) \cdot  \mathscr{C}(x)  
- \partial^\mu[ \mathscr{A}_\mu(x) \cdot \mathscr{C}(x)] .
\end{align}
In the Landau gauge $\partial^\mu \mathscr{A}_\mu(x)=0$, therefore, the BRST transformation of $\phi(x)$ is reduced to the total derivative and the spacetime average of $\phi(x)$ is BRST invariant, i.e., 
$\bm{\delta}_{\rm B} \int d^4x \phi(x)=0$.

\subsection{Total Yang-Mills action}

A purpose of this paper is to obtain an equivalent theory which does not involve the gluon fields $\mathscr{A}_\mu$ by integrating them out.  
The GF+FP term in the Lorentz gauge $\mathscr{L}_{\rm GF+FP}$ (\ref{GF+FP2}) includes the linear and quadratic terms in the gluon field.  

The pure Yang-Mills Lagrangian is further rewritten as follows. 
For this purpose, we adjust the parameters to eliminate the  cubic and quartic interaction terms in the field $\mathscr{A}_\mu$.
The cubic interaction term in (\ref{YM2}) is  eliminated by choosing 
\begin{align}
 \rho = \sigma^{-1} .
  \label{relation}
\end{align}
Moreover, by choosing the parameters as 
\begin{align}
  \sigma_\phi = {3(\sigma^2-1)g^2 \over N_c}, \quad
  \sigma_G = {4(1-\sigma^2)g^2 \over N_c}, \quad
  \sigma_\varphi =  (3/2)(\sigma^2-1)g^2 , \quad
  \sigma_V =  2(1-\sigma^2)g^2  ,
  \label{sigmaparameter}
\end{align}
the quartic gluon interaction terms in (\ref{Aux}) are also eliminated. 
Then $\sigma$ is  only one undetermined parameter at this stage and it must be larger than one, i.e., $\sigma^2>1$, to realize the condensation as shown later. 

Due to the analysis \cite{FK78,Fukuda77,CS83}, the triple gluon interaction provides the attractive force, whereas the quartic gluon interaction can give the repulsive force among the gluons.  Therefore, for the gluon pair condensation to occur, the attractive force coming from three-point vertex dominates the repulsive force from four-point vertex to yield the net attractive interaction between a pair of gluons.  

In our treatment, the original triple gluon vertex is completely eliminated by introducing the antisymmetric tensor $B_{\mu\nu}$ and choosing $\rho = \sigma^{-1}$. Therefore, the net attractive interaction is reflected in the relation $\sigma^2>1$. In fact, if this is not the case, i.e., repulsive interaction $1-\sigma^2>0$, the gluon pair condensation does not occur as will be shown later. 
\par
Therefore, the pure Yang-Mills Lagrangian $\mathscr{L}_{\rm YM}$ (\ref{YM2}) becomes at most quadratic in the gluon field and  reduces to
\footnote{ 
If we want to leave two independent parameters $\rho$ and $\sigma$, we have only to add
\begin{align}
   +{1-\rho \sigma \over 2} g (\partial_\mu \mathscr{A}_\nu - \partial_\nu \mathscr{A}_\mu) \cdot (\mathscr{A}^\mu \times \mathscr{A}^\nu) .
\end{align}
Therefore, we can choose an interesting option $\rho=\sigma$, i.e., 
$B_{\mu\nu}$ is the Hodge dual of $\mathscr{F}_{\mu\nu}$, 
although the cubic gluon interaction term remains in this case. 
} 
\begin{align}
  \mathscr{L}_{\text{YM}} 
  =&  {1-\rho^2 \over 4} (\partial_\mu \mathscr{A}_\nu - \partial_\nu \mathscr{A}_\mu)^2 
+{1 \over 4} B_{\mu\nu} \cdot B^{\mu\nu} 
\nonumber\\& 
- {i \over 2} B^{\mu\nu} \cdot {}^*[\rho (\partial_\mu \mathscr{A}_\nu - \partial_\nu \mathscr{A}_\mu) + \sigma g (\mathscr{A}_\mu \times \mathscr{A}_\nu)]
\nonumber\\& 
   + {\sigma_\phi \over 2} \phi^2 + {\sigma_\phi \over 2} \phi (\mathscr{A}_\mu \cdot \mathscr{A}^\mu)
 % \nonumber\\ &
  + {\sigma_G \over 2} G_{\mu\nu} G^{\mu\nu} + {\sigma_G \over 2} G_{\mu\nu} S^{\mu\nu}
  \nonumber\\ &
  + {\sigma_\varphi \over 2} \varphi \cdot \varphi + {\sigma_\varphi \over 2} \varphi \cdot (\mathscr{A}_\mu \star \mathscr{A}^\mu)
%  \nonumber\\ &
  + {\sigma_V \over 2} V_{\mu\nu} \cdot V^{\mu\nu} + {\sigma_V \over 2} V_{\mu\nu} \cdot T^{\mu\nu} .
  \label{YM4}
\end{align}

%In the total Lagrangian, $\mathscr{J}^\mu{}^A$ is modified due to the existence of the Nakanishi-Lautrup field $B^A$ and FP ghost $C^A$ and antighost field $\bar{C}^A$.  
Moreover, the GF+FP term is reduced to 
\begin{align}
  \mathscr{L}_{\rm GF+FP} 
=&  {1 \over 2\lambda}(\partial_\mu \mathscr{A}_\mu)^2
+  \bar{\mathscr{C}} \cdot \partial_\mu \partial_\mu \mathscr{C}
+ (1-\xi)g    \mathscr{A}_\mu \cdot (\partial_\mu \bar{\mathscr{C}} \times \mathscr{C})
\nonumber\\&
- \xi g  \mathscr{A}_\mu \cdot (\bar{\mathscr{C}} \times \partial_\mu \mathscr{C})
- {1 \over 2}\lambda \xi (1-\xi) g^2 ( \mathscr{C} \times \bar{\mathscr{C}}) \cdot ( \mathscr{C} \times \bar{\mathscr{C}})
\nonumber\\&
 + {\sigma_{FP} \over 2} \phi_{FP}^2 
+ {\sigma_{FP}' \over 2} \varphi_{FP} \cdot \varphi_{FP}   
+ \sigma_{FP} \phi_{FP}  \bar{\mathscr{C}} \cdot \mathscr{C} 
 + \sigma_{FP}' \varphi_{FP} \cdot (\bar{\mathscr{C}} \star \mathscr{C}) .
 \label{GF+FP3}
\end{align}
Here the four-ghost interaction terms in the GF+FP term are eliminated by choosing 
\begin{align}
   \sigma_{FP} = \lambda \xi (1-\xi) g^2 , \quad 
 \sigma_{FP}' = \lambda \xi (1-\xi) g^2  .
\end{align}
Finally, we obtain the total Lagrangin which is at most quadratic in the gluon field $\mathscr{A}_\mu$:
\begin{subequations}
\begin{align}
  \mathscr{L}_{\text{YM}}^{\rm tot} 
  =& 
 {1 \over 4} B_{\mu\nu} \cdot B^{\mu\nu} 
  + {\sigma_\phi \over 2} \phi^2  
  + {\sigma_G \over 2} G_{\mu\nu} G^{\mu\nu}   
  + {\sigma_\varphi \over 2} \varphi \cdot \varphi  
  + {\sigma_V \over 2} V_{\mu\nu} \cdot V^{\mu\nu}  
\nonumber\\& 
+ {\sigma_{FP} \over 2} \phi_{FP}^2 
+ {\sigma_{FP}' \over 2} \varphi_{FP} \cdot \varphi_{FP}   
\nonumber\\& 
  +  \bar{\mathscr{C}} \cdot \partial_\mu \partial^\mu \mathscr{C}
+ \sigma_{FP} \phi_{FP}  \bar{\mathscr{C}} \cdot \mathscr{C} 
 + \sigma_{FP}' \varphi_{FP} \cdot (\bar{\mathscr{C}} \star \mathscr{C})
\nonumber\\& 
+ {1 \over 2} \mathscr{A}_\mu^A \mathscr{K}^{AB}{}^{\mu\nu} \mathscr{A}_\nu^B  + \mathscr{A}_\mu \cdot \mathscr{J}^\mu ,
\end{align}
where we have defined $\mathscr{K}^{AB}_{\mu\nu}$ and $\mathscr{J}_\mu^A$ by  
\begin{align}
 \mathscr{K}^{AB}_{\mu\nu} :=& \delta^{AB} \left[ -(1-\rho^2) (\partial^2 \delta_{\mu\nu} - \partial_\mu \partial_\nu) 
 -  \lambda^{-1} \partial_\mu \partial_\nu  \right]
  \nonumber\\ &
  -i g \sigma f^{ABC} {}^*B_{\mu\nu}^C 
  +  \sigma_\phi \delta^{AB} \delta_{\mu\nu} \phi 
  +   \sigma_\varphi d^{ABC} \delta_{\mu\nu} \varphi^C 
  \nonumber\\ &
  +\sigma_G \delta^{AB} \left( G_{\mu\nu}-{1 \over 4}\delta_{\mu\nu}G_\rho{}^\rho \right)
  +\sigma_V d^{ABC} \left( V_{\mu\nu}^C-{1 \over 4}\delta_{\mu\nu}V^C_\rho{}^\rho \right)   ,
\\ 
  \mathscr{J}_\mu^A :=&  J_\mu^A  - i \rho \partial_\nu {}^*B_{\mu\nu}^{A} + (1-\xi)g   (\partial_\mu \bar{\mathscr{C}} \times \mathscr{C})^A
- \xi g  (\bar{\mathscr{C}} \times \partial_\mu \mathscr{C})^A .
\end{align}
  \label{YM5}
\end{subequations}
Here  $J_\mu^A$ is the source for the gluon field $\mathscr{A}_\mu^A$. Note that $\mathscr{K}^{AB}_{\mu\nu}$ is at most linear in the auxiliary field and $\mathscr{J}_\mu^A$ contains the bilinear term in the ghost and antighost fields. 
 The source term for the auxiliary field will be introduced later to avoid the complication.

Thus, we can integrate out the gluon field $\mathscr{A}_\mu$ exactly:
\begin{align}
  S_{\text{EYM}} 
  =& \int d^4x \Big\{
+{1 \over 4} B_{\mu\nu} \cdot B^{\mu\nu} 
  + {\sigma_\phi \over 2} \phi^2  
  + {\sigma_G \over 2} G_{\mu\nu} G^{\mu\nu}   
  + {\sigma_\varphi \over 2} \varphi \cdot \varphi  
  + {\sigma_V \over 2} V_{\mu\nu} \cdot V^{\mu\nu}  
\nonumber\\& 
+ {\sigma_{FP} \over 2} \phi_{FP}^2 
+ {\sigma_{FP}' \over 2} \varphi_{FP} \cdot \varphi_{FP}   
\nonumber\\& 
  +  \bar{\mathscr{C}} \cdot \partial_\mu \partial^\mu \mathscr{C}
+ \sigma_{FP} \phi_{FP}  \bar{\mathscr{C}} \cdot \mathscr{C} 
 + \sigma_{FP}' \varphi_{FP} \cdot (\bar{\mathscr{C}} \star \mathscr{C})
\nonumber\\& 
- {1 \over 2} \mathscr{J}_\mu^A [\mathscr{K}^{AB}{}^{\mu\nu}]^{-1} \mathscr{J}_\nu^B   \Big\} 
+{1 \over 2} \ln {\rm Det} [\mathscr{K}^{AB}{}^{\mu\nu}] .
  \label{YM7}
\end{align}
When $\lambda=0$, $\sigma_{F}=0=\sigma_{FP}'$ and the ghost self-interaction term disappears. 
Therefore, we do not need to introduce $\phi_{FP}$ and $\varphi_{FP}^A$ in (\ref{YM7}) which decouple  from the theory. 
Integrating out the gluon field $\mathscr{A}_\mu^A$ and FP ghost fields $\mathscr{C}^A, \bar{\mathscr{C}}^A$, we can obtain the effective theory  written in terms of five kinds of auxiliary fields 
$B_{\mu\nu}^A$, $\phi$, $\varphi^A$, $G_{\mu\nu}$, $V_{\mu\nu}^A$. 
The $\lambda\not=0$ case will be considered later.

Some remarks are in order. 

1. The resulting theory has one undetermined parameter $\sigma$.  In the classical level, any value of $\sigma$ reproduces the original Yang-Mills theory, as is trivial from the above construction. 
However, in the quantum level, this is not the case.  The renormalization  urges the parameter $\sigma$ to run according to the renormalization scale, as will be discussed in Appendix A. 

2. For a special choice $\sigma^2=1$, we have $\rho^2 =\sigma^2=1$ and hence the coefficients vanish $\sigma_\Phi=0$, which implies that all the auxiliary fields  
$\phi, G_{\mu\nu}, \varphi, V_{\mu\nu}$
except for $B_{\mu\nu}$ 
decouple from the theory. 
Therefore, this case reproduces the field strength formulation \cite{Halpern77} of the Yang-Mills theory, which was also called the deformation of the topological BF theory \cite{BFYM,KondoII} 
with the Lagrangian:  
\begin{align}
  \mathscr{L}_{\text{BF}} 
%  =  {1 \over 4} B_{\mu\nu} \cdot B^{\mu\nu} 
%- {i \over 2} B^{\mu\nu} \cdot {}^*[\partial_\mu \mathscr{A}_\nu - \partial_\nu \mathscr{A}_\mu + g \mathscr{A}_\nu \times \mathscr{A}_\nu]
  = 
 {1 \over 4} B_{\mu\nu} \cdot B^{\mu\nu} 
- {i \over 2} B^{\mu\nu} \cdot {}^*\mathscr{F}_{\mu\nu}  .
  \label{YM6}
\end{align}
(The signature of $\rho$ and $\sigma$ can be absorbed by the redefinition of the fields $\mathscr{A}_\mu$ and $B_{\mu\nu}$.)
  When $\sigma^2 \not=1$, therefore, we have the scalar field $\phi, \varphi$ and the symmetric tensor fields $G_{\mu\nu}, V_{\mu\nu}$ in addition to the antisymmetric tensor field $B_{\mu\nu}$.

3. It is in principle possible to introduce the auxiliary field 
$\phi_T$ corresponding to the square 
${1 \over 2} (\mathscr{A}_{\text{phys}}^i)^2$ 
of the {\it transverse} gluon modes 
$\mathscr{A}_{\text{phys}}^i$
(i.e., $\partial_i \mathscr{A}_{\text{phys}}^i(x)=0$) 
as
\begin{align}
  1 = \int \mathcal{D}\phi_T \exp \left\{ - \int d^4x 
{\sigma_\phi \over 2} \left( \phi_T + {1 \over 2} (\mathscr{A}_{\text{phys}}^i)^2
\right)^2 
\right\} ,
\end{align}
where $\mathscr{A}_{\text{phys}}^i$ is  both 
BRST invariant, 
$\bm{\delta}_{\rm B} \mathscr{A}_{\text{phys}}^i(x)=0$,
 and gauge invariant, 
$\bm{\delta}_{\rm \omega} \mathscr{A}_{\text{phys}}^i(x)=0$.  
Then the auxiliary field $\phi_T$ introduced in this way becomes BRST invariant, 
$\bm{\delta}_{\rm B} \phi_T(x)=0$, and gauge invariant, 
$\bm{\delta}_{\rm \omega} \phi_T(x)=0$. 
For example, such a transverse mode is written as
\begin{align}
  \mathscr{A}_{\text{phys}}^i(x) = \left( \delta^{ij} - \partial^i {1 \over \Delta} \partial^j  \right) \Phi^j(x) , \quad 
\Phi^j = \Phi^j_{(0)} + g \Phi^j_{(1)} + g^2 \Phi^j_{(2)} + \cdots ,
\label{Tmode}
\end{align}
where $\Phi^j_{(n)}$ can be determined 
order by order of the coupling constant $g$.
  The first two terms are  
$
 \Phi^j_{(0)}=\mathscr{A}^j  
$
and 
$
 \Phi^j_{(1)} = [v_{(1)}, \mathscr{A}^j] + {1 \over 2}[\partial^j v_{(1)}, v_{(1)}]  
$
where 
$
 v_{(1)} := {1 \over \Delta} \partial_j \mathscr{A}^j 
$,
see section 5 and Appendix A of \cite{LM94} for more details. 
Needless to say,  the transverse mode (\ref{Tmode}) reduces in the Abelian limit to the well-known expression: 
\begin{align}
  \bm{A}^T(x) = \left( 1 - \nabla {1 \over \Delta} \nabla \cdot \right) \bm{A}(x) 
= \bm{A}(x^0,\bm{x}) + \partial_i^x \int d^3 \bm{y} {1 \over 4\pi|\bm{x}-\bm{y}|} \partial_j^y A^j(x^0,\bm{y})  .
\label{AT}
\end{align}
However, in this case, the gluon self-interaction can not be eliminated, since $\mathscr{A}_{\text{phys}}^i$ is a non-local and furthermore non-linear quantity written in terms of $\mathscr{A}_\mu$ in the non-Abelian gauge theory. (Even the all order expression of (\ref{Tmode}) is not yet known.)  
Therefore, the strategy adopted in this paper is difficult to be applied to this case which is more transparent from the physical point of view. 

%\newpage
\subsection{Another reformulation}

We can introduce another antisymmetric tensor field $B_{\mu\nu}^A{}'$ by
\begin{align}
  B_{\mu\nu}^A  \rightarrow B_{\mu\nu}^A{}' := 
B_{\mu\nu}^A-i \rho {}^* (\partial_\mu \mathscr{A}_\nu - \partial_\nu \mathscr{A}_\mu)^A  ,
\end{align}
which is equivalent to the identification as an auxiliary field:
\begin{align}
    B_{\mu\nu}{}' \leftrightarrow \sigma g {}^* (\mathscr{A}_\mu \times \mathscr{A}_\nu) .
\end{align}
Then the Lagrangian is modified as 
\begin{align}
  \mathscr{L}_{\text{YM}} 
  =&  
\frac{1}{4} B_{\mu\nu}{}' \cdot B^{\mu\nu}{}' - {i \over 2} B^{\mu\nu}{}' \cdot {}^*[\sigma g (\mathscr{A}_\mu \times \mathscr{A}_\nu)] 
\nonumber\\ &
+ {1 \over 4} (\partial_\mu \mathscr{A}_\nu - \partial_\nu \mathscr{A}_\mu)^2 
\nonumber\\
 &  +{1 \over 2} g (\partial_\mu \mathscr{A}_\nu - \partial_\nu \mathscr{A}_\mu) \cdot (\mathscr{A}^\mu \times \mathscr{A}^\nu) 
\nonumber\\
 &+{1-\sigma^2 \over 4} g^2 (\mathscr{A}_\mu \times \mathscr{A}_\nu) \cdot (\mathscr{A}^\mu \times \mathscr{A}^\nu) .
 \label{YM2'}
\end{align}
By this redefinition of the antisymmetric tensor field, the original cross term $- {i \over 2} B^{\mu\nu} \cdot \rho  {}^*(\partial_\mu \mathscr{A}_\nu - \partial_\nu \mathscr{A}_\mu)$ has disappeared. 
The price is that the three-gluon interactions remain  in this case. 
The four-gluon interactions can be eliminated by applying the same procedure  as that given above. 
Thus we obtain the modified Yang-Mills Lagrangian%
\footnote{
This Lagrangian is obtained from (\ref{YM4}) by setting $\rho=0$ and adding the three-gluon interaction term.
Therefore, the modified total Lagrangian  is also obtained from (\ref{YM5}) by the same procedure.
Note that setting $\rho=0$ is equivalent to eliminating the cross propagator between $A_\mu$ and $B_{\alpha\beta}$ in the diagrams.
}
\begin{align}
  \mathscr{L}_{\text{YM}} 
  =&  {1 \over 4} (\partial_\mu \mathscr{A}_\nu - \partial_\nu \mathscr{A}_\mu)^2 
 + \frac{1}{4} B_{\mu\nu}{}' \cdot B^{\mu\nu}{}' - {i \over 2} B^{\mu\nu}{}' \cdot {}^*[\sigma g (\mathscr{A}_\mu \times \mathscr{A}_\nu)] 
\nonumber\\
 &  +{1 \over 2} g (\partial_\mu \mathscr{A}_\nu - \partial_\nu \mathscr{A}_\mu) \cdot (\mathscr{A}^\mu \times \mathscr{A}^\nu) 
\nonumber\\& 
   + {\sigma_\phi \over 2} \phi^2 + {\sigma_\phi \over 2} \phi (\mathscr{A}_\mu \cdot \mathscr{A}^\mu)
 % \nonumber\\ &
  + {\sigma_G \over 2} G_{\mu\nu} G^{\mu\nu} + {\sigma_G \over 2} G_{\mu\nu} S^{\mu\nu}
  \nonumber\\ &
  + {\sigma_\varphi \over 2} \varphi \cdot \varphi + {\sigma_\varphi \over 2} \varphi \cdot (\mathscr{A}_\mu \star \mathscr{A}^\mu)
%  \nonumber\\ &
  + {\sigma_V \over 2} V_{\mu\nu} \cdot V^{\mu\nu} + {\sigma_V \over 2} V_{\mu\nu} \cdot T^{\mu\nu} .
  \label{YM4'}
\end{align}
In order to treat the $\rho=\sigma$ case, anyway, the three-gluon interaction can not be eliminated. 
The absence of the four-gluon interactions eliminates the two-loop diagrams in the Schwinger-Dyson equation for the full gluon propagator.  Although we must deal simultaneously with a number of  auxiliary field propagators and the cross propagator between the gluon and the antisymmetric tensor field, all the diagrams are of the one-loop type and there is no ambiguity encountered in treating the two-loop diagrams. If we set $\sigma=0$, furthermore, the antisymmetric tensor field $B_{\mu\nu}{}'$ decouples from the theory. 
Thus this reformulation helps to simplify the Schwinger-Dyson equation for the full gluon propagator, since  the absence of cross propagator reduces the number of propagators to be investigated. 
The details will be given elsewhere \cite{KMS04}.

%%%%%%%%%%%%%%%%%%%%%%%%%%%%%%%%%%%%%%%%%%%%%%%%%%%%%%%%%%%%%%%%%%%%%
%%%%%%%%%%%%%%%%%%%%%%%%%%%%%%%%%%%%%%%%%%%%%%%%%%%%%%%%%%%%%%%%%%%%%
%%%%%%%%%%%%%%%%%%%%%%%%%%%%%%%%%%%%%%%%%%%%%%%%%%%%%%%%%%%%%%%%%%%%%
%%%%%%%%%%%%%%%%%%%%%%%%%%%%%%%%%%%%%%%%%%%%%%%%%%%%%%%%%%%%%%%%%%%%%
%%%%%%%%%%%%%%%%%%%%%%%%%%%%%%%%%%%%%%%%%%%%%%%%%%%%%%%%%%%%%%%%%%%%%
\section{Setting up the $1/N_c$ expansion}

\begin{figure}[htbp]
\begin{center}
\includegraphics[height=5cm]{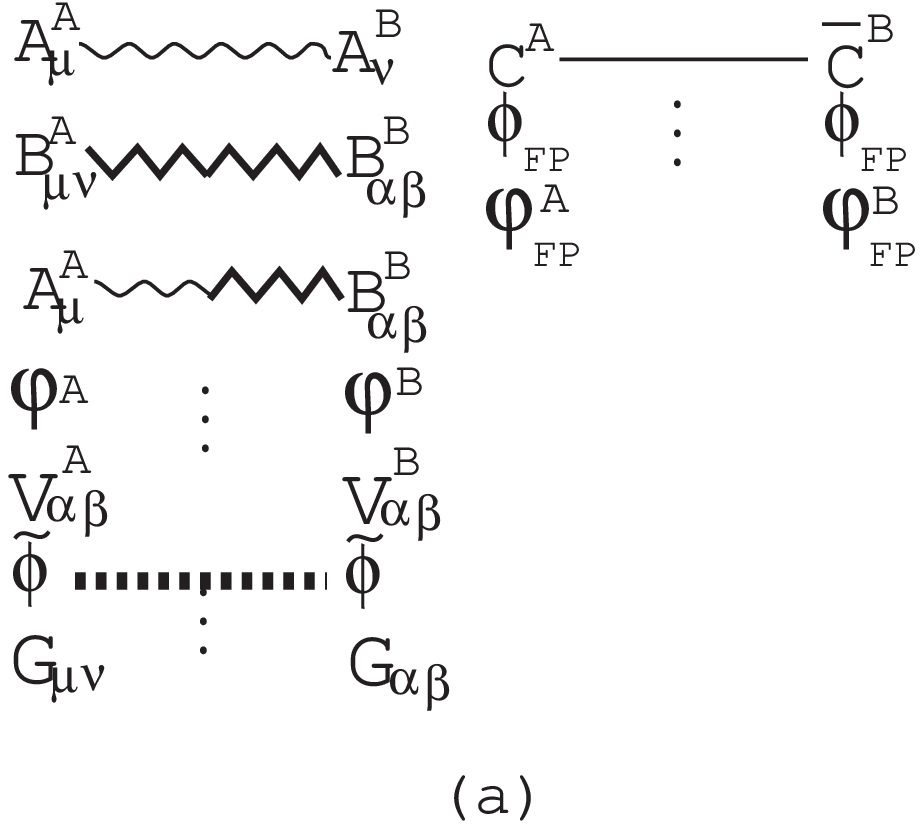}
\includegraphics[height=5cm]{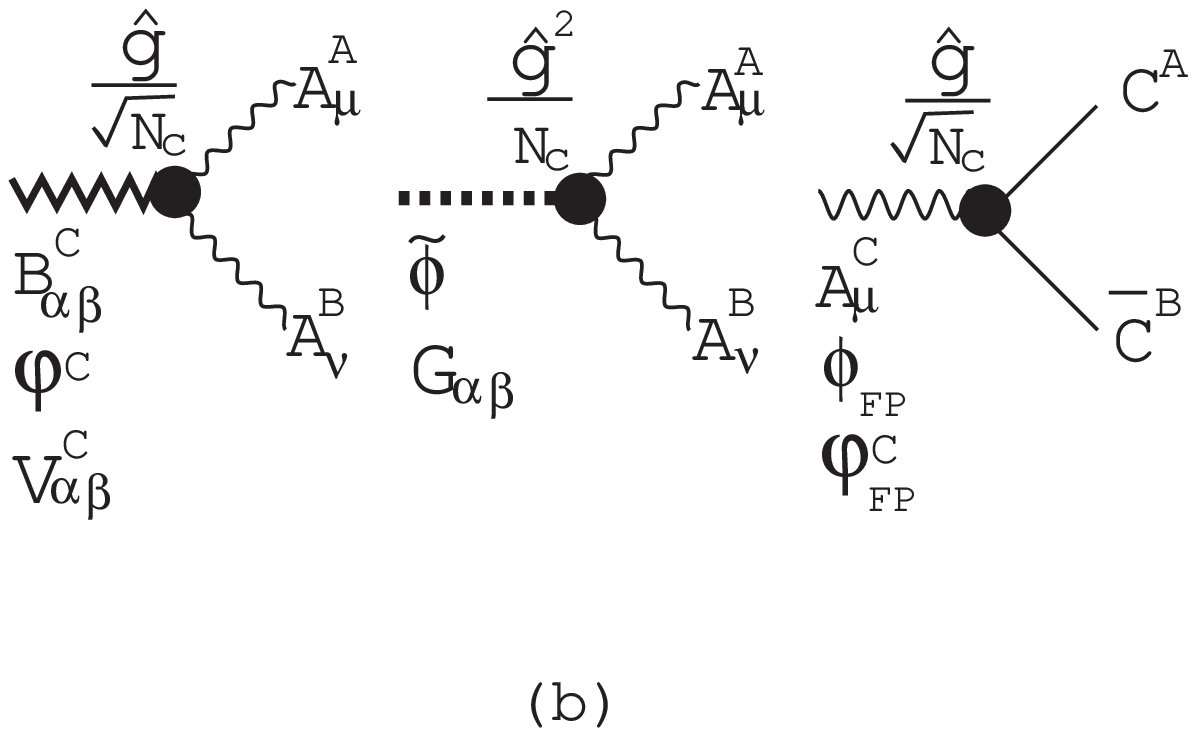}
\caption{Feynman rules of the $1/N_c$ expansion with the 't Hooft coupling $\hat{g}^2$.
(a)~Propagators for the gluon $\mathscr{A}_\mu^A$ (wavy line), two set of auxiliary fields $B_{\mu\nu}^A, \varphi^A, V_{\mu\nu}^A$ (zigzag line)  and  and  $\tilde\phi, G_{\mu\nu}$ (broken line),
and FP ghost, antighost $\mathscr{C}^A, \bar{\mathscr{C}}^A$, the auxiliary fields $\phi_{FP}, \varphi_{FP}$ (solid line)
 (b) Vertices for two gluons $\mathscr{A}_\mu^A$ and one auxiliary fields from a set $B_{\mu\nu}^A, \varphi^A, V_{\mu\nu}^A$ or a set $\tilde\phi, G_{\mu\nu}$, and vertices for ghost-antighost and one gluon field or auxiliary fields $\phi_{FP}, \varphi_{FP}$. 
There are no four-point interaction vertices. 
}
\label{fig:FeynmanRule}
\end{center}
\end{figure}

In the previous section, we have rewritten the Yang-Mills theory into an equivalent form given by (\ref{YM5}).
We set up the $1/N_c$ expansion for this theory as follows. 
 For this theory, the graphical representation of the Feynman rules  are given in Fig.~\ref{fig:FeynmanRule}.
Here the propagators and the vertices are easily read off from the total Lagrangian (\ref{YM5}). 
Note that there is a cross-propagator between $A_\mu$ and $B_{\alpha\beta}$ for $\rho\not=0$, i.e., except for the another formulation.

 From the viewpoint of the $1/N_c$  expansion, the field $G_{\mu\nu}$ behave as $\tilde\phi$  from the identity 
\begin{equation}
  \delta^{AB}\delta^{AB}=N_c^2-1 . 
\end{equation}
Other fields $\varphi^A, V_{\mu\nu}^A$ behave as $B_{\mu\nu}^A$ in the large $N_c$, since the identity holds 
\begin{equation}
  d^{ACD}d^{BCD}= \delta^{AB} (N_c^2-4)/N_c , 
\end{equation}
which is similar for large $N_c$ to the relationship
\begin{equation}
  f^{ACD}f^{BCD}=\delta^{AB} N_c  .
\end{equation}
This is also the case for $\phi_{FP}$ and $\varphi_{FP}^A$ and for $\mathscr{C}^A$ and $\bar{\mathscr{C}}^A$. 
Therefore, the leading contributions in the condensed vacuum $\phi_0\not=0$ come from the interaction of $\mathscr{A}_\mu^A$ with $B_{\mu\nu}^A, \varphi^A, V_{\mu\nu}^A, \mathscr{C}^A, \bar{\mathscr{C}}^A$ in the Landau gauge $\lambda=0$ and also from $\phi_{FP}$ and $\varphi_{FP}^A$ in the non-Landau gauge $\lambda\not=0$. 
%%%%%%%%%%%%%%%%%%%%%%%%%%%%%%%%%%%%%%%%%%%%%%%%%%%%%%%%%%%%

%%%%%%%%%%%%%%%%%%%%%%%%%%%%%%%%%%%%%%%%%%%%%%%%%%%%%%%%%%%%

In the following, we attempt to sum the leading order diagrams in the $1/N_c$ expansion. 
We have rewritten the original Yang-Mills theory into an equivalent theory with the generating functional:
\begin{align}
  Z[J] := \int \mathcal{D}\mathscr{\phi}\mathcal{D}B_{\mu\nu}^A \mathcal{D}G_{\mu\nu} \mathcal{D}\mathscr{\varphi}^A \mathcal{D}V_{\mu\nu}^A
\mathcal{D}\mathscr{C}^A \mathcal{D}\bar{\mathscr{C}}^A 
 [\mathcal{D}\phi_{FP} \mathcal{D}\varphi_{FP}^A] 
\exp \left\{ -S_{E} \right\} ,
\end{align}
where the source terms are included for the original fields and those for the auxiliary fields are omitted here (but they will be introduced later).

First, for simplicity,  we consider only a sector containing $\phi$ and $\mathscr{A}_\mu^A$.
The generating functional is given by
\begin{align}
  Z[J] :=& \int   \mathcal{D}\mathscr{\phi} \exp \left\{ -S_{E}[\phi;J] \right\} , 
\label{Z1}
\\
 S_{E}[\phi;J] :=&  \int_{x}  {N_c^2 \over 6(\sigma^2-1)\hat{g}^2} \phi(x)^2 
 +{N_c^2-1 \over 2} {\rm tr} \ln [\mathscr{K}{}^{\mu\nu}] 
\nonumber\\ &
 + \int_{x} {-1 \over 2} J_\mu^A [\mathscr{K}^{\mu\nu}]^{-1} J_\nu^A
,
\label{S_E1}
\end{align}
where the gluon field $\mathscr{A}_\mu^A$ has been integrated out and its contribution is replaced by the source term $J_\mu^A$,
and we have introduced the notation $\int_{x}:=\int d^4x$ and 
\begin{align}
\mathscr{K}{}_{\mu\nu}[\phi](x) := -(1-\rho^2) (\partial^2 \delta_{\mu\nu} - \partial_\mu \partial_\nu) 
 -  \lambda^{-1} \partial_\mu \partial_\nu 
  +   \delta_{\mu\nu} \phi .
\end{align}
Here we have made the replacement
\begin{align}
  g^2 \rightarrow \hat{g}^2/N_c 
\end{align}
 with the 't~Hooft coupling $\hat{g}^2$ being fixed, and
 we have redefined the auxiliary field  
 $\phi \rightarrow \phi/\sigma_\phi$, 
 which corresponds to the identification:
\begin{align}
  \phi(x) \leftrightarrow  - \sigma_\phi \frac{1}{2} \mathscr{A}_\mu^A(x) \mathscr{A}_\mu^A(x) 
%  = - \sigma_\phi {\rm tr} [\mathscr{A}_\mu(x) \mathscr{A}_\mu(x) ] 
  = - 3(\sigma^2-1){\hat{g}^2 \over N_c^2}  {\rm tr} [\mathscr{A}_\mu(x) \mathscr{A}_\mu(x) ] . 
  \label{normal}
\end{align}
The field $\phi$ is a single component color singlet field, whereas each gluon field 
$\mathscr{A}_\mu:=\mathscr{A}_\mu^A T^A$ is an adjoint field with $N_c^2-1$ components. 
The normalization factor in (\ref{normal}) provides the natural normalization, since the field $\phi(x)$ produces a color singlet state 
${\rm tr} [\mathscr{A}_\mu(x) \mathscr{A}_\mu(x)]$ with unit amplitude.  
Note that the first two terms in $S_{E}[\phi;J]$ are proportional to $N_c^2$ for large $N_c$. 
The effective action $S_E[\phi;J]$ has no hidden factors of $N_c$ in this sector. 
Thus the overall factor of $N_c$ does imply that the large $N_c$ limit is the same as the semiclassical limit of an effective theory of color singlet field $\phi$ (glueball) at least in this sector. 

 For the current theory to be meaningful in the limit $N_c \rightarrow \infty$, therefore, the expansion around $N_c=\infty$ in the integral of (\ref{Z1})  can be performed by  means of the steepest descent method (or saddle point method). 
The stationary point $\phi_0(x)$ of $\phi(x)$ is obtained from the saddle-point equation:
\begin{align}
 {1 \over 3(\sigma^2-1)\hat{g}^2} \phi_0(x)
+{1 \over 2} {\rm tr} (\mathscr{K}^{-1}_{\mu\mu}[\phi_0(x)])  = 0 .
\label{gap0}
\end{align}
This is nothing but the gap equation. The regularized form is given in  the next section under the dimensional regularization.  This equation has two branches, the trivial branch $\phi=0$ and the nontrivial branch $\phi\not=0$ for arbitrary values of parameters. 
The nontrivial solution of (\ref{gap0}) agrees with the stationary point $\phi_0$ away from the origin of the effective potential. 
It turns out that the nontrivial solution is preferred, since it lowers the energy of the system, as shown in the next section.
If the vacuum is translation invariant, the stationary value $\phi_0(x)$ does not depend on $x$. 

Note that $(\mathscr{A}_\mu^A)^2$ is not a gauge invariant quantity. 
Nevertheless, it is possible to extract the gauge-invariant part from  $(\mathscr{A}_\mu^A)^2$ which is written as the sum of a gauge-invariant part and a non-invariant part, which implies the decomposition 
$\phi(x)=\phi_{inv}(x)+\phi_{non}(x)$ is possible \cite{Kondo03c} where $\phi_{inv}$ is the gauge invariant part and $\phi_{non}$ is the gauge non-invariant, but BRST exact part.
Then the vacuum expectation value (VEV) of the gauge non-invariant part vanishes 
$
 \langle \phi_{non}(x) \rangle =0
$,
and consequently only the VEV of the gauge invariant part gives a non-vanishing value
$
 \langle \phi(x) \rangle = \langle \phi_{inv}(x) \rangle \not=0 
$,
  which is identified with $\phi_0$ \cite{Kondo03c}.

\begin{figure}[htbp]
\begin{center}
\includegraphics[height=2cm]{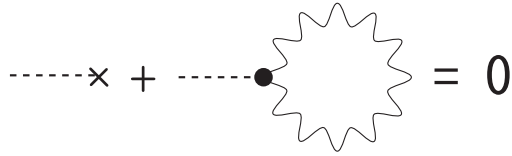}
\caption{Elimination of the tadpole contribution. The broken line denotes the auxiliary field $\tilde\phi$ and the wavy line the gluon field $\mathscr{A}_\mu^A$.}
\label{fig:tadpole_dilaton}
\end{center}
\end{figure}

Substituting $\phi(x)=\phi_0+{1 \over N_c}\tilde{\phi}(x)$ into the action $S_{E}$, and expanding the action around the stationary point $\phi_0$ in powers of $\tilde{\phi}(x)$, we obtain (up to field-independent constants)
\begin{align}
 S_{E}' =& N_c \int_{x} \ \left\{ {1 \over 3(\sigma^2-1)\hat{g}^2} \phi_0  
+{1 \over 2} {\rm tr} (\mathscr{K}^{-1}_{\mu\mu}[\phi_0])   \right\} 
 \tilde{\phi}(x) 
\nonumber\\ &
+ \int_{x} \  {1 \over 2} \left\{ {1 \over 3(\sigma^2-1)\hat{g}^2}  - {1 \over 2} {\rm tr} (\mathscr{K}^{-1}_{\mu\rho}[\phi_0]
\mathscr{K}^{-1}_{\rho \mu}[\phi_0]) 
\right\} \tilde{\phi}^2(x) 
\nonumber\\ &
 + \int_{x}  {-1 \over 2} J_\mu^A [\mathscr{K}^{\mu\nu}[\phi_0]]^{-1} J_\nu^A
+  O(1/N_c) .
\label{expan1}
\end{align}
%where 
%$
% \mathscr{K}^{-1}_{\mu\nu}[p;\phi_0] = [(1-\rho^2)p^2+\phi_0]^{-1}P_{\mu\nu}^T(p) + \lambda [(1-\rho^2)p^2+\lambda \phi_0]^{-1} P_{\mu\nu}^L(p) 
%$
%with 
%$P^T_{\mu\nu}(p):=\delta_{\mu\nu}-p_\mu p_\nu/p^2$
%and 
%$P^L_{\mu\nu}(p):= p_\mu p_\nu/p^2$.
The first term of (\ref{expan1}) shows that the saddle-point equation (\ref{gap0}) is equivalent to requiring cancellation of the tadpole term. See Fig.~\ref{fig:tadpole_dilaton} for the graphical representation. 
%%%%%%%%%%%%%%%%%%%%%%%%%%%%%%%%%%%%%%%%%%%%%%%%%%%%%%%%%%%%
%%%%%%%%%%%%%%%%%%%%%%%%%%%%%%%%%%%%%%%%%%%%%%%%%%%%%%%%%%%%

The  connected $n$-point correlation function  is obtained by differentiating $n$ times the generating function $W[J]:=\ln Z[J]$ with respect to the source. 
For example, the gluon propagator is obtained from 
\begin{align}
   \langle \mathscr{A}_\mu^A(x) \mathscr{A}_\nu^B(y) \rangle_{conn}
   = {\delta \over \delta J_\mu^A(x)} {\delta \over \delta J_\nu^B(y)} \ln Z[J]  .
\label{2point}
\end{align}
In the $1/N_c$ expansion, 
the gluon propagator $\mathscr{D}_{\mu\nu}^{AB}(k^2)$ in momentum space is read off from the source term in (\ref{S_E1}) as 
\begin{align}
  \mathscr{D}_{\mu\nu}^{AB}(k^2) &:= \delta^{AB} \mathscr{K}^{-1}_{\mu\nu}[k;\phi_0]
%\nonumber\\&
%= \delta^{AB} \left\{ [(1-\rho^2)k^2+\phi_0]^{-1}P_{\mu\nu}^T(k) + \lambda [(1-\rho^2)k^2+\lambda \phi_0]^{-1} P_{\mu\nu}^L(k) \right\} ,
\nonumber\\&
= \delta^{AB} \left\{ {1 \over (1-\rho^2)k^2+\phi_0}P_{\mu\nu}^T(k) + {\tilde\lambda \over (1-\rho^2)k^2+\tilde\lambda \phi_0} P_{\mu\nu}^L(k) \right\} ,
\end{align}
where we have introduced the rescaled gauge parameter $\tilde\lambda:=(1-\rho^2)\lambda$, the transverse projection
$P^T_{\mu\nu}(p):=\delta_{\mu\nu}-p_\mu p_\nu/p^2$
and the longitudinal one 
$P^L_{\mu\nu}(p):= p_\mu p_\nu/p^2$. 
This result  shows that the gluon acquires the dynamical mass given by
\begin{align}
 M^2:={\phi_0 \over 1-\rho^2} 
%={\sigma^2 \over \sigma^2-1} \phi_0 
.
\end{align}

\par
It is important to remark that in the first term in the right-hand side (RHS) of (\ref{expan1}) or (\ref{gap0}), there exists the logarithmic divergence coming from the second term 
${1 \over 2} {\rm tr} (\mathscr{K}^{-1}_{\mu\mu}[\phi_0])$, which can be absorbed into the renormalization of the first term 
${1 \over 3(\sigma^2-1)\hat{g}^2}$. 
The same divergence appears in the second term  ${1 \over 2} {\rm tr} (\mathscr{K}^{-1}_{\mu\rho}[\phi_0]
\mathscr{K}^{-1}_{\rho \mu}[\phi_0])$ in the RHS of (\ref{expan1}).
Therefore, $\sigma$ should be regarded as the renormalized one $\sigma_R$ in what follows. 
See Appendix A for explicit verification of this fact based on the dimensional regularization in the modified minimal subtraction scheme ($\overline{\text{MS}}$).%
\footnote{The quadratic divergence $\Lambda^2 \tilde\phi$ appearing in the term
${1 \over 2} {\rm tr} [\mathscr{K}^{-1}_{\mu\mu}[\phi_0]]\tilde{\phi}$ with the ultraviolet momentum cutoff $\Lambda$ 
can be removed by introducing the counterterm of the form, $C \tilde\phi$ (the same form as the source term $J_\phi \tilde\phi$ for $\tilde\phi$).
In the dimensional regularization, such quadratic divergence is hidden and does not appear. 
For this purpose,  we can introduce the auxiliary field $\phi$ in (\ref{aux-phi}) by replacing $\phi$ with $\phi+C$  for the translation invariant measure $\mathcal{D}\phi$.
Then we have additional terms, 
$\frac{1}{2}C \mathscr{A}_\mu^2 + C \phi$ which is equal to 
$\frac{1}{2}C \mathscr{A}_\mu^2 + C \tilde\phi$ apart form an irrelevant constant $C\phi_0$.
The redundant term $\frac{1}{2}C \mathscr{A}_\mu^2$ here is removed by introducing the counterterm of  Curci-Ferrari type, $-\frac{1}{2}C \mathscr{A}_\mu^2$, (in the Landau gauge $\lambda=0$) in the bare Lagrangian. 
In other words, removing the quadratic divergence $\Lambda^2 \tilde\phi$ is equivalent to introducing the counterterm 
$\frac{1}{2}\Lambda^2 \mathscr{A}_\mu^2$. 
It should be remarked that the non-zero gluon mass is generated from the term $\frac{1}{2}\phi_0 \mathscr{A}_\mu^2$ by further removing the logarithmic divergence, irrespective of the quadratic divergence. 

}

\begin{figure}[htbp]
\begin{center}
\includegraphics[height=2cm]{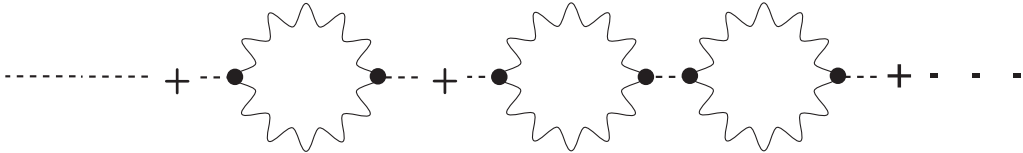}
\caption{The propagator (\ref{propa-phi}) of the auxiliary field $\tilde\phi$ as the summation of the infinite geometric series of the massive gluon-loop chain.}
\label{fig:dilaton_propagator}
\end{center}
\end{figure}

The connected correlation function 
$\langle \phi(x); \phi(y) \rangle_{conn}$
shows the factorization 
\begin{align}
 \langle \phi(x); \phi(y) \rangle_{conn} := 
  \langle \phi(x) \phi(y) \rangle -  \langle \phi(x) \rangle  \langle \phi(y) \rangle ={1 \over N_c^2} \langle \tilde\phi(x) \tilde\phi(y) \rangle .
  \label{con-con}
\end{align}
The auxiliary field  propagator $\mathscr{D}_{\tilde\phi\tilde\phi}(k^2)$ is read off from the second integrand of   (\ref{expan1}) based on the formula (\ref{2point}) after introducing the source term $J_{\phi} \tilde{\phi}$:
\begin{align}
  \mathscr{D}_{\tilde\phi\tilde\phi}(k^2) 
= \left\{   {1 \over 3(\sigma^2-1)\hat{g}^2} -{1 \over 2} \int {d^4p \over (2\pi)^4}  \mathscr{K}^{-1}_{\mu\nu}[p;\phi_0]  \mathscr{K}^{-1}_{\nu\mu}[p+k;\phi_0] 
\right\}^{-1} . 
\label{propa-phi}
\end{align}
Thus the auxiliary field  propagator $\mathscr{D}_{\tilde\phi\tilde\phi}(k^2)$ has the well-defined  
  UV behavior due to the cancellation of the UV divergence between two terms.
It turns out that the propagator  $\mathscr{D}_{\tilde\phi\tilde\phi}(k^2)$  corresponds to the summation of the infinite geometric series of the {\it massive} gluon-loop chain, see Fig.~\ref{fig:dilaton_propagator}.
%%%%%%%%%%%%%%%%%%%%%%%%%%%%%%%%%%%%%%%%%%%%%%%%%%%%%%%%%%%%
%%%%%%%%%%%%%%%%%%%%%%%%%%%%%%%%%%%%%%%%%%%%%%%%%%%%%%%%%%%%

In the small momentum $k^2$ region, the propagator  for $\tilde\lambda=0$ or $1$ is calculated in the form: 
\begin{align}
  \mathscr{D}_{\tilde\phi\tilde\phi}(k^2) =   {192\pi^2 (3+\tilde\lambda)^{-1}(1-\rho^2)^2 M^2 \over k^2+6M^2}       ,
\label{glueball}
\end{align}
by using the gap equation (\ref{gap0}) and the result of Appendix A.
The connected correlation function  
$\langle \phi(x); \phi(y) \rangle_{conn}$
decays exponentially for large $|x-y|$,
 since the right-hand-side of (\ref{con-con}) is equal to the Fourier transform of (\ref{glueball}). 
Here it should be understood that the contributions from the gauge non-invariant part vanish in the final expression. In this way, we can obtain the mass of the lightest scalar glueball $0^{++}$. 
The ratio of the glueball mass $0^{++}$ to the gluon mass is given by 
\begin{align}
  M(0^{++})/M = \sqrt{6} \cong 2.45 ,
\end{align}
in the leading order. 
This ratio is gauge independent, i.e., independent of the gauge fixing parameter $\lambda$ and it does not depend on the parameter $\sigma$. 
This value  is similar to the prediction of the potential model of Cornwall and Soni \cite{CS83,HLW01}: 
$
 M(0^{++})/M \cong 2.3
$. 
If one adopts the gluon mass $M = 600 \sim 700 \text{MeV}$  as  suggested from the numerical simulation on a lattice \cite{LRG02}, this ratio reproduces the expected value of the glueball mass \cite{MP99}
$
  M(0^{++}) = 1.5 \sim 1.7 \ \text{GeV}
$.
Heavier gluon mass around $1 \ \text{GeV}$ is also suggested, see phenomenological analyses \cite{Field02}, lattice simulation \cite{LSWP99} and the analytical study \cite{KK95}.
\par
Next, we include the antisymmetric tensor field $B_{\mu\nu}^A$ and consider a sector consisting of $\phi$, $B_{\mu\nu}^A$ and $\mathscr{A}_\mu^A$.  The generating functional is given by 
\begin{align}
  Z[J] :=& \int   \mathcal{D}\mathscr{\phi}  \mathcal{D}B_{\mu\nu}^A \exp \left\{ -S_{E} \right\} , 
\nonumber\\
 S_{E} :=& \int_{x} \  {1 \over 4} B_{\mu\nu}(x) \cdot B^{\mu\nu}(x) 
 + 
\int_{x} \   {N_c^2 \over 6(\sigma^2-1)\hat{g}^2} \left( \phi_0+{1 \over N_c}\tilde{\phi}(x) \right)^2 
\nonumber\\ &
+ \int_{x} {-1 \over 2} \mathscr{J}_\mu^A [\mathscr{K}^{AB}_{\mu\nu}]^{-1}  \mathscr{J}_\nu^B 
 +{1 \over 2} \ln \det [\mathscr{K}^{AB}_{\mu\nu}] ,
\label{S_E3}
\end{align}
where 
\begin{align}
\mathscr{K}^{AB}_{\mu\nu} :=& \mathscr{K}_{\mu\nu}[\phi_0]\delta^{AB}
  +   {1 \over N_c}\delta^{AB} \delta_{\mu\nu} \tilde{\phi} 
    -i  {\hat{g} \over \sqrt{N_c}} \sigma  f^{ABC} {}^*B_{\mu\nu}^C ,
  \nonumber\\ 
  \mathscr{J}_\mu^A :=&  J_\mu^A  
-i \rho \partial_\nu {}^*B_{\mu\nu}^{A} .
\label{K2}
\end{align}
The action has the $1/N_c$ expansion around $\phi_0$ (up to field-independent constants):
\begin{align}
 S_{E}' =&   N_c  \int_{x} \  \left[{1 \over 3(\sigma^2-1)\hat{g}^2} \phi_0
+ {1 \over 2} {\rm tr} (\mathscr{K}^{-1}_{\mu\mu}[\phi_0])
  \right]  \tilde{\phi}
 \nonumber\\ &
+ \int_{x} \ {1 \over 2} \left[ {1 \over 3(\sigma^2-1)\hat{g}^2} 
- {1 \over 2} {\rm tr} \left( 
\mathscr{K}^{-1}_{\mu\rho}[\phi_0]
\mathscr{K}^{-1}_{\rho \mu}[\phi_0]
 \right) \right]  \tilde{\phi}^2 
\nonumber\\ &
+ \int_{x} \Big\{ \Big[ {1 \over 4}  I_{\mu\nu,\alpha\beta}
- {1 \over 4}\hat{g}^2 \sigma^2 {\rm tr} \left( 
  I_{\mu\nu,\rho_1 \sigma_1} 
\mathscr{K}^{-1}_{\sigma_1 \sigma_2}[\phi_0] 
  I_{\sigma_2\rho_2 ,\alpha\beta}
\mathscr{K}^{-1}_{\rho_2 \rho_1}[\phi_0]
%(  {}^*B_{\mu\sigma} \cdot  {}^*B_{\sigma\nu}) 
\right)  \Big] {}^*B^{\mu\nu} \cdot {}^*B^{\alpha\beta}
\nonumber\\ & \quad\quad\quad\quad
  + {1 \over 2}\rho^2 \partial_\lambda {}^*B^{\mu\lambda}{}^{A} 
   \mathscr{K}_{\mu\nu}^{-1}[\phi_0] \delta^{AB}
   \partial_\tau {}^*B^{\nu\tau}{}^{B} 
   \Big\} 
\nonumber\\ &
 + \int_{x}  {-1 \over 2} J_\mu^A [\mathscr{K}^{\mu\nu}[\phi_0]]^{-1} J_\nu^A
+  O(1/N_c) ,
\label{leading2}
\end{align}
where 
we have defined 
$I_{\mu\nu,\alpha\beta} := {1 \over 2}(\delta_{\mu\alpha}\delta_{\nu\beta}-\delta_{\mu\beta}\delta_{\nu\alpha})$,
and
we have used the identities: 
$\delta^{AB}\delta^{AB}=N_c^2-1$ and  
$f^{ACD}f^{BCD}=N_c \delta^{AB}$. 
 In (\ref{leading2}), 
a contribution coming from the term 
$\mathscr{J}_\mu^A$
 is simplified as
\begin{align}
  {1 \over 2}\rho^2 \partial_\lambda {}^*B^{\mu\lambda}{}^{A} 
   \mathscr{K}_{\mu\nu}^{-1}[\phi_0] \delta^{AB}
   \partial_\tau {}^*B^{\nu\tau}{}^{B} 
 =  {1 \over 2}
   {\rho^2 \over -(1-\rho^2)\partial^2+\phi_0} 
   (\partial_\nu {}^*B^{\mu\nu}{}^{A})^2  ,
\end{align}
where the gauge-parameter dependent part in   $\mathscr{K}_{\mu\nu}^{-1}[\phi_0]$ disappears due to 
the antisymmetric property of $B_{\mu\nu}$.
Therefore, the kinetic term for $B_{\mu\nu}$ is generated and the auxiliary field becomes dynamical in the leading order of $1/N_c$ expansion irrespective of the magnitude of the 't Hooft coupling $\hat{g}$, as far as $\phi_0 \not=0$.%
\footnote{This phenomenon has already been claimed to occur in \cite{Kondo00,KI02} based on the perturbation expansion in the coupling constant to a certain order.  Therefore, it was unreliable to extend the result to the infrared region where the coupling constant becomes quite large. 
}

\begin{figure}[htbp]
\begin{center}
\includegraphics[height=5cm]{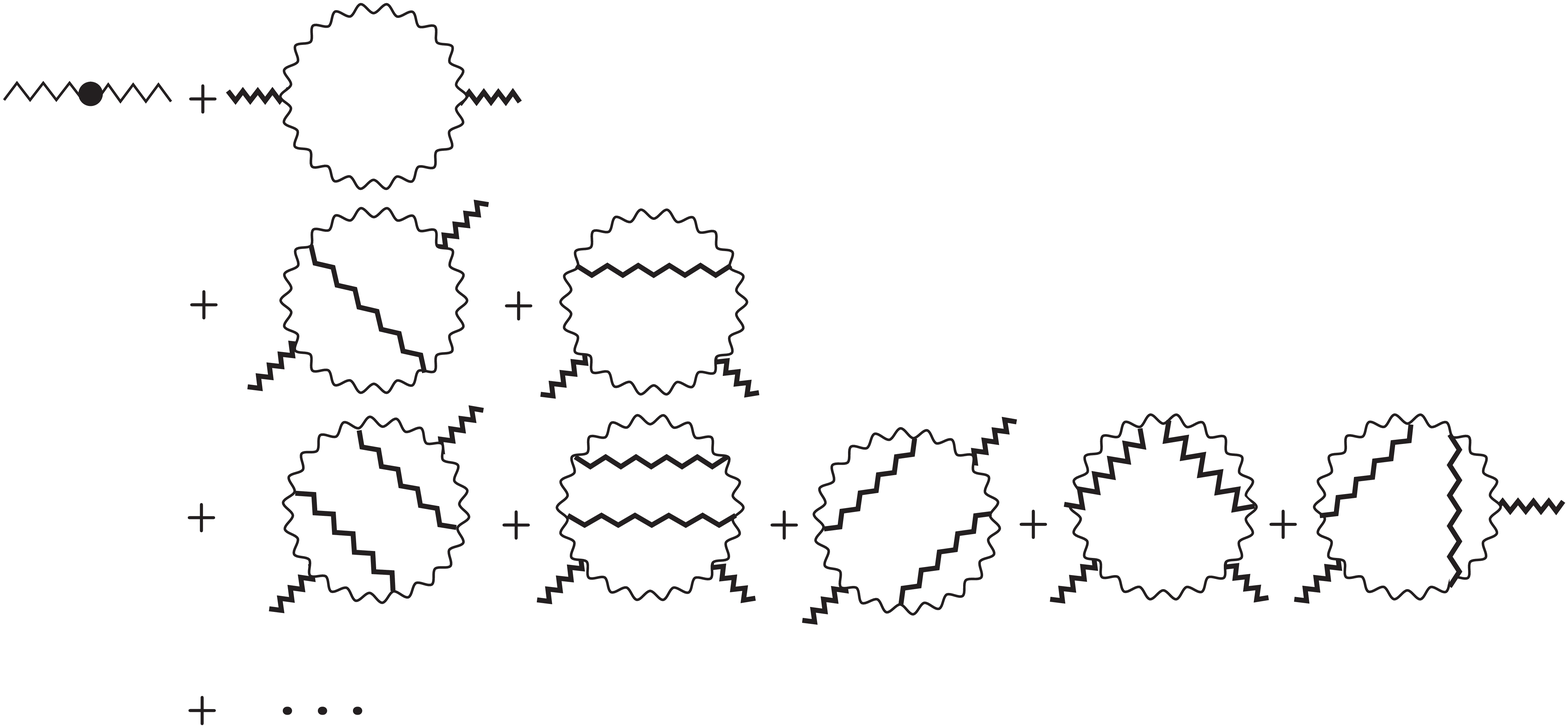}
\caption{Planar diagrams in the leading order for the inverse propagator of the auxiliary tensor  field $B_{\mu\nu}$ where the zig-zag line denotes the $B_{\mu\nu}$ field and the wavy line the gluon field.}
\label{fig:B_planar}
\end{center}
\end{figure}

In the action, the terms with more than three auxiliary fields are suppressed in $1/N_c$. Hence the leading contribution  is exhausted by a few terms. 
However, such simplicity is apprarent and is immediately lost, as soon as we proceed to obtain the propagator. 
In fact, the auxiliary tensor field propagator $\mathscr{D}^{AB}_{B_{\mu\nu},B_{\alpha\beta}}(k^2)$ in the leading order consists of  all the planar diagrams with two external lines corresponding to $B_{\mu\nu}$ and $B_{\alpha\beta}$ (see Fig.~\ref{fig:B_planar}) where the bare propagator is given  by%
\footnote{
We have used the integration by parts  
\begin{align}
 \int_{x} (\partial_\nu {}^*B^{\mu\nu}{}^{A})^2 
&= \int_{x} {1 \over 4}(\epsilon^{\mu\nu\rho\sigma}\partial_\nu B_{\rho\sigma}^{A})^2
\nonumber\\
&= \int_{x} {1 \over 2}(-B_{\mu\nu}^{A} \partial^2 B_{\mu\nu}^{A} - B_{\mu\nu}^{A} \partial_\nu \partial_\lambda B_{\lambda\mu}^{A} + B_{\nu\mu}^{A} \partial_\mu \partial_\lambda B_{\lambda\nu}^{A}) .
\end{align}
For simplicity, we have imposed the {\it gauge-fixing} condition 
$
 \partial_\lambda B_{\lambda\nu} = 0 
$, 
although this is not an indispensable condition, see e.g., \cite{KI02} for more details.
}
\begin{align}
%  \mathscr{D}^{AB}_{B_{\mu\nu},B_{\alpha\beta}}{}^{-1}(k^2) 
 \mathscr{D}^{(0)}{}^{AB}_{B_{\mu\nu},B_{\alpha\beta}}(k^2)
 :=& \left\{ {1 \over 2} I_{\mu\nu,\alpha\beta} \delta^{AB}
 +    {\rho^2 \over 2(1-\rho^2)} 
 {k^2 \over k^2+M^2} I_{\mu\nu,\alpha\beta} \delta^{AB}
\right\}^{-1} ,
\label{propa-B0}
\end{align}
and the vertex is read off from the action (\ref{S_E3}) and (\ref{K2}). 
Such planar diagrams are obtained from the gluon one-loop diagram with a number of external fields $B$ by contracting the external lines two by two.  In each step, a factor $N$ is increased without increasing the order of $g$. 
All the diagrams in Fig.~\ref{fig:B_planar} are the same order in the $1/N_c$ expansion as can be seen by making use of the double line representation. 

The similar remark is also applied to the propagator of the auxiliary field $\tilde\phi$. 
The inverse propagator of the auxiliary field $\tilde\phi$ in the leading order is the sum of the similar diagrams to Fig.~\ref{fig:B_planar} where two external $B$ lines are replaced by the  $\tilde\phi$.  This is because the graphs with internal  $\tilde\phi$-lines give  higher oder contributions in $1/N_c$. 
%%%%%%%%%%%%%%%%%%%%%%%%%%%%%%%%%%%%%%%%%%%%%%%%%%%%%%%%%%%%
%%%%%%%%%%%%%%%%%%%%%%%%%%%%%%%%%%%%%%%%%%%%%%%%%%%%%%%%%%%%

\begin{figure}[htbp]
\begin{center}
\includegraphics[height=2cm]{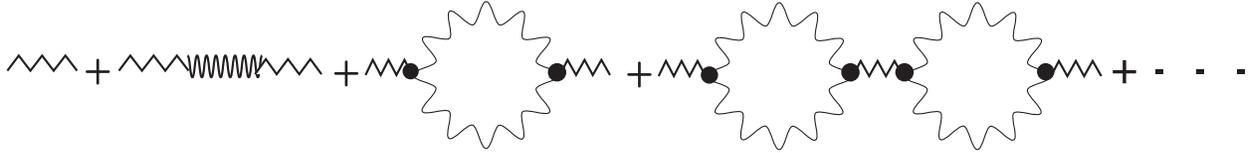}
\caption{The propagator (\ref{propa-B}) of the auxiliary tensor  field $B_{\mu\nu}$  as the summation of the infinite geometric series of the massive gluon-loop chain.
Here the sum of the first two propagators denotes the propagator (\ref{propa-B0}).}
\label{fig:B_propagator}
\end{center}
\end{figure}

%The effective action $S_{E}'$ has the part:
%$
% {1 \over 2}\int{d^4k \over (2\pi)^4} {}^*B^{\mu\nu}{}^A (k)
%\mathscr{D}^{AB}_{\mu\nu,\alpha\beta}{}^{-1}(k^2)
% {}^*B^{\alpha\beta}{}^B(-k)  
%$
Thus, it is still impossible to sum up all the planar diagrams in question to have the closed form for the propagator in the leading of the $1/N_c$ expansion. 
Here we obtain the auxiliary tensor field propagator, the inverse of which  is truncated up to $O(\hat{g}^2)$  as above
\begin{align}
%  \mathscr{D}^{AB}_{B_{\mu\nu},B_{\alpha\beta}}{}^{-1}(k^2) 
 \mathscr{D}^{AB}_{B_{\mu\nu},B_{\alpha\beta}}(k^2)
 :=& \left\{ {1 \over 2} I_{\mu\nu,\alpha\beta} \delta^{AB}
 +    {\rho^2 \over 2(1-\rho^2)} 
 {k^2 \over k^2+M^2} I_{\mu\nu,\alpha\beta} \delta^{AB}
+  \Pi_{\mu\nu,\alpha\beta}^{AB}(k) 
\right\}^{-1} ,
\label{propa-B}
\end{align}
where  $\Pi_{\mu\nu,\alpha\beta}^{AB}(k)$ is the vacuum polarization tensor in the lowest order given by
\begin{align}
\Pi_{\mu\nu,\alpha\beta}^{AB}(k) 
:=&   \delta^{AB} 
%\left\{ 
%{1 \over 2} I_{\mu\nu,\alpha\beta} +
{1 \over 2}  \hat{g}^2 \sigma^2  \int {d^4p \over (2\pi)^4} 
  I_{\mu\nu,\rho_1 \sigma_1} 
\mathscr{K}^{-1}_{\sigma_1 \sigma_2}[p;\phi_0]
  I_{\alpha\beta,\rho_2 \sigma_2}
 \mathscr{K}^{-1}_{\rho_2 \rho_1}[p+k;\phi_0]
.
% \right\} .
\end{align}
The explicit form of the vacuum polarization tensor $\Pi_{\mu\nu,\alpha\beta}^{AB}(k)$ is given  based on  $\overline{\text{MS}}$ in  Appendix B. 
The obtained propagator corresponds to the summation of the infinite geometric series of the {\it massive} gluon-loop chain, see  Fig.~\ref{fig:B_propagator} for the graphical representation.
This propagator obtained by the partial resummation of the leading order graphs is called the {\it truncated propagator}  hearafter.
It is possible to resum the diagrams based on the Schwinger-Dyson equation approach which will be given elsewhere \cite{KMS04}. 
%%%%%%%%%%%%%%%%%%%%%%%%%%%%%%%%%%%%%%%%%%%%%%%%%%%%%%%%%%%%
%%%%%%%%%%%%%%%%%%%%%%%%%%%%%%%%%%%%%%%%%%%%%%%%%%%%%%%%%%%%

\par
In the small momentum $k^2$ region, the vacuum polarization tensor for $B_{\mu\nu}$ reads%
\footnote{
 The divergent term is absorbed into the renormalization of the tensor field $B_{\mu\nu}$ through the renormalization constant $Z_B$ defined by  
$B_{\mu\nu}=Z_B^{1/2} B_{\mu\nu}^R$, just as in the case of Maximal Abelian gauge  \cite{KS01,Kondo01}. 
}
\begin{align}
\Pi_{\mu\nu,\alpha\beta}^{AB}(k) 
 =&  -
  \delta^{AB}  { \hat{g}^2    \over 16\pi^2} {\sigma^2 \over (1-\rho^2)^2}
  I_{\mu\nu,\alpha\beta} \Biggr[  - {1+\lambda \over 2} \frac{1}{\bar{\epsilon}}
+  f_0 + f_1 {k^2 \over M^2} + f_2 {k^4 \over M^4}   \Biggr] 
\nonumber\\&
 + O\left(  {k^6 \over M^6} \right) ,
\end{align}
where  
%$M^2:=\phi_0/(1-\rho^2)$, 
$
 f_0 =   {1 \over 2} \ln {M^2 \over \mu^2} - {1 \over 4} ,
%- {1 \over 2}\gamma_E + {1 \over 2}\ln 4\pi , 
\quad f_1 = {11 \over 180} ,  
\quad f_ 2 =   {4 \over 105}
$
for the Landau gauge $\lambda=0$
and 
$
 f_0 =    \ln {M^2 \over \mu^2} ,
%- {1 \over 2}\gamma_E + {1 \over 2}\ln 4\pi , 
\quad f_1 = {19 \over 120} ,  
\quad f_ 2 =   {29 \over 1680}
$
for the Feynman gauge $\lambda=1$.
Thus, the truncated propagator 
$\mathscr{D}^{AB}_{B_{\mu\nu},B_{\alpha\beta}}(k^2)$  of the auxiliary field $B_{\mu\nu}$ 
has the IR asymptotic behavior 
\begin{align}
  \mathscr{D}^{AB}_{B_{\mu\nu},B_{\alpha\beta}}(k^2) 
=   \delta^{AB} I_{\mu\nu,\alpha\beta}
% 16\pi^2 (1-\rho^2)^2 \sigma^{-2} f_1^{-1} 
 \left[ {1 \over 2(\sigma^2-1)}- {\hat{g}^2 \over 16\pi^2}{\sigma^2 \over (1-\rho^2)^2} f_1 \right]^{-1} 
   {M^2 \over m^2 + k^2  + O(k^4)} ,
\label{B-ir}
\end{align}
where 
%$m^2:=f_0 f_1^{-1}M^2 -8\pi^2 \hat{g}^{-2}(1-\rho^2)^2 \sigma^{-2} f_1^{-1} M^2$.
\begin{align}
  m^2 := M^2 \left[ {1 \over 2(\sigma^2-1)}- {\hat{g}^2 \over 16\pi^2}{\sigma^2 \over (1-\rho^2)^2} f_1 \right]^{-1} 
  \left[ \frac{1}{2}-{\hat{g}^2 \over 16\pi^2}{\sigma^2 \over (1-\rho^2)^2} f_0 \right] . 
\end{align}

\par
In the large  momentum $k^2$ region, on the other hand, the truncated  propagator reads 
\begin{align}
  \mathscr{D}^{AB}_{B_{\mu\nu},B_{\alpha\beta}}(k^2) 
=  \delta^{AB}  I_{\mu\nu,\alpha\beta} 16\pi^2 (1-\rho^2)^2 \sigma^{-2} &
\Biggr\{ 
  {k^2 \over M^2} \left( {1 \over 6}\ln {k^2 \over M^2} -{5 \over 18} \right) 
+ O( \ln k^2)
\Biggr\}^{-1} .
\label{B-uv}
\end{align}
If such a high energy behavior as (\ref{B-uv}) is preserved after the inclusion of all the planar diagrams,  our theory of  gluons with non-vanishing  effective  mass is expected to be renormalizable order by order in the $1/N_c$ expansion. 
This is in sharp contrast to the naive massive Yang-Mills theory which is known to be nonrenormalizable in weak-coupling perturbation theory.  
More arguments on this point will be given in a subsequent paper. 

Other fields can be treated in the similar way. 
In fact, after an appropriate rescaling of other auxiliary fields as $\phi$, 
the total action reads 
\begin{align}
  S_{\text{EYM}} 
  =& \int_{x} \Big\{
 {1 \over 4} B_{\mu\nu} \cdot B^{\mu\nu} 
  + {\sigma_\phi^{-1} \over 2} \phi^2  
  + {\sigma_G^{-1} \over 2} G_{\mu\nu} G^{\mu\nu}   
\nonumber\\& 
  + {1 \over 2} \varphi \cdot \varphi  
  + {1 \over 2} V_{\mu\nu} \cdot V^{\mu\nu}  
+ {1 \over 2} \phi_{FP}^2 
+ {1 \over 2} \varphi_{FP} \cdot \varphi_{FP}   
\nonumber\\& 
  +  \bar{\mathscr{C}} \cdot \partial_\mu \partial^\mu \mathscr{C}
+  \sqrt{\sigma_{FP}}\phi_{FP}  \bar{\mathscr{C}} \cdot \mathscr{C} 
 +  \sqrt{\sigma_{FP}'} \varphi_{FP} \cdot (\bar{\mathscr{C}} \star \mathscr{C})
%\nonumber\\& 
%+ {1 \over 2} \mathscr{A}_\mu^A \mathscr{K}^{AB}{}^{\mu\nu} \mathscr{A}_\nu^B  + \mathscr{A}_\mu \cdot \mathscr{J}^\mu ,
\nonumber\\& 
- {1 \over 2} \mathscr{J}_\mu^A [\mathscr{K}^{AB}_{\mu\nu}]^{-1} \mathscr{J}_\nu^B   \Big\} 
+{1 \over 2} \ln {\rm Det} [\mathscr{K}^{AB}_{\mu\nu}] ,
\end{align}
where  
\begin{align}
 \mathscr{K}^{AB}_{\mu\nu} :=& \mathscr{K}_{\mu\nu}[\phi_0]   \delta^{AB} 
  -i  {\hat{g}  \over \sqrt{N_c}}  \sigma f^{ABC} {}^*B_{\mu\nu}^C 
  +  N_c^{-1}  \delta^{AB} \delta_{\mu\nu}  \tilde\phi 
  +   \sqrt{\sigma_\varphi}  d^{ABC} \delta_{\mu\nu} \varphi^C 
  \nonumber\\ &
  + \delta^{AB} \left( G_{\mu\nu}-{1 \over 4}\delta_{\mu\nu}G_\rho{}^\rho \right)
  + \sqrt{\sigma_V} d^{ABC} \left( V_{\mu\nu}^C-{1 \over 4}\delta_{\mu\nu}V^C_\rho{}^\rho \right)   ,
  \nonumber\\ 
  \mathscr{J}_\mu^A :=&  J_\mu^A  - i \rho \partial_\nu {}^*B_{\mu}{}^{\nu}{}^{A} + (1-\xi) {\hat{g}  \over \sqrt{N_c}}   (\partial_\mu \bar{\mathscr{C}} \times \mathscr{C})^A
- \xi {\hat{g}  \over \sqrt{N_c}}  (\bar{\mathscr{C}} \times \partial_\mu \mathscr{C})^A .
\end{align}

%%%%%%%%%%%%%%%%%%%%%%%%%%%%%%%%%%%%%%%%%%%%%%%%%%%%%%%%%%%%%%%%%%%%%
%%%%%%%%%%%%%%%%%%%%%%%%%%%%%%%%%%%%%%%%%%%%%%%%%%%%%%%%%%%%%%%%%%%%%
%%%%%%%%%%%%%%%%%%%%%%%%%%%%%%%%%%%%%%%%%%%%%%%%%%%%%%%%%%%%%%%%%%%%%
%%%%%%%%%%%%%%%%%%%%%%%%%%%%%%%%%%%%%%%%%%%%%%%%%%%%%%%%%%%%%%%%%%%%%
%%%%%%%%%%%%%%%%%%%%%%%%%%%%%%%%%%%%%%%%%%%%%%%%%%%%%%%%%%%%%%%%%%%%%

\section{Estimate of gluon mass and vacuum condensate}

Now, we examine based on the effective potential whether the vacuum condensate of mass dimension two is realized or not. 
All the auxiliary fields $B_{\mu\nu}^A$, $\phi$, $\varphi^A$, $G_{\mu\nu}$, $V_{\mu\nu}^A$ have the canonical mass dimension 2.  
Therefore, various vacuum condensates can contribute to lower the energy of the system.  However, if the color rotation invariance and Lorentz rotation invariance are required, we find
$
 \langle B_{\mu\nu}^A \rangle = \langle \varphi^A \rangle =\langle G_{\mu\nu} \rangle =\langle V_{\mu\nu}^A \rangle = 0 
$.
Only a possibility examined in this paper is 
$
 \langle \phi \rangle \not= 0 
$.
This vacuum condensate corresponds to
$
 \langle (\mathscr{A}_\mu^A)^2  \rangle \not= 0 
$
which
breaks neither color symmetry nor Lorentz symmetry. 
In what follows, we treat the vacuum condensate of mass dimension 2 alone.%
\footnote{
This does not exclude the dimension four condensate of the type:
$\langle B_{\mu\nu}^A B_{\mu\nu}^A\rangle$, $\langle \varphi^A \varphi^A \rangle$, $\langle G_{\mu\nu} G_{\mu\nu} \rangle$, $\langle V_{\mu\nu}^A V_{\mu\nu}^A\rangle \not= 0 $. 
}

The minimum value of the effective potential should not depend on the gauge fixing parameter.  Therefore, we calculate only the $\lambda=0$ case.  In this case, the BRST-invariant composite operator reduces to $\mathscr{A}_\mu^2$.

\begin{figure}[htbp]
\begin{center}
\includegraphics[height=5cm]{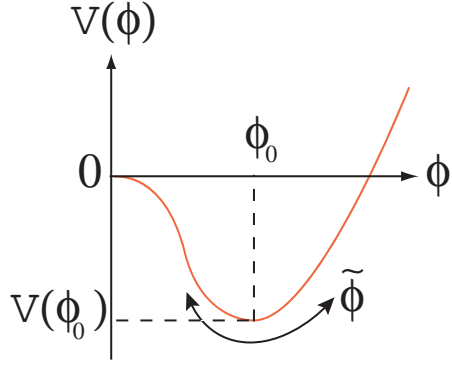}
\caption{The effective potential for the auxiliary field $\phi$. 
}
\label{fig:non-trivial_vev}
\end{center}
\end{figure}

The effective potential is obtained from%
\footnote{We use the same letters to denote the field obtained after the Legendre transformation.
}
\begin{align}
V = {\sigma_\phi \over 2} \phi^2 + {1 \over 2} \ln \det [-(1-\rho^2) \delta^{AB}(\partial^2 \delta_{\mu\nu} - \partial_\mu \partial_\nu) 
 - \delta^{AB}{1 \over \lambda} \partial_\mu \partial_\nu 
 +   \delta^{AB} \delta_{\mu\nu} \sigma_\phi \phi ] .
\end{align}
The scalar part of the effective potential is obtained by resumming the gluon one-loop contribution as
\begin{align}
V(\phi) =& {\sigma_\phi \over 2} \phi^2 + {N_c^2-1 \over 2} \int {d^4p \over (2\pi)^4} \ln \det \left[ 
 (1-\rho^2) (p^2 \delta^{\mu\nu} - p^\mu p^\nu) 
 + {1 \over \lambda} p^\mu p^\nu
 +  \delta^{\mu\nu} \sigma_\phi \phi 
\right]  
  \nonumber\\  
=& {\sigma_\phi \over 2} \phi^2 + {N_c^2-1 \over 2} \int {d^4p \over (2\pi)^4} \ln  \left\{ \left[  (1-\rho^2) p^2 + \sigma_\phi \phi \right]^3  
\left[ p^2 + \lambda \sigma_\phi \phi \right]/\lambda \right\} .
\end{align}
At $\lambda=0$ (up to $\phi$-independent constant), we obtain 
\begin{align}
V(\phi) =& {\sigma_\phi \over 2} \phi^2  + {3(N_c^2-1) \over 2} \int {d^4p \over (2\pi)^4} \ln  \left[ 
 1   + {{\sigma_\phi \over 1-\rho^2} \phi \over  p^2}
\right]  
  \nonumber\\  
=& {\sigma_\phi \over 2} \phi^2  + {3(N_c^2-1) \over 64\pi^2} \left({\sigma_\phi \over 1-\rho^2} \phi \right)^2  \ln  {{\sigma_\phi \over 1-\rho^2} \phi \over 4\pi\mu^2} ,
\end{align}
where $V(0)=0$. 
The first derivative reads
\begin{align}
V'(\phi) =& \phi \left[  \sigma_\phi  + {3(N_c^2-1) \over 64\pi^2}\left({\sigma_\phi \over 1-\rho^2} \right)^2 + {3(N_c^2-1) \over 32\pi^2} \left({\sigma_\phi \over 1-\rho^2} \right)^2  \ln  {{\sigma_\phi \over 1-\rho^2} \phi \over 4\pi\mu^2} \right] .
\end{align}
The stationary condition $V'(\phi)=0$ has two solutions, $\phi=0$ or $\phi=\phi_0\not=0$. See Fig.~\ref{fig:non-trivial_vev}. 
Thus the absolute minimum of the potential is given at $\phi=\phi_0$
%\begin{align}
%  \phi = {1-\rho^2 \over g^2}4\pi\mu^2 \exp \left\{ -\left[ {\sigma_1 \over 2} + {3(N_c^2-1) \over 64\pi^2}\left({g^2 \over 1-\rho^2} \right)^2 \right] {32\pi^2 \over 3(N_c^2-1)} \left({1-\rho^2 \over g^2} \right)^2 \right\} 
\label{DVeq}
%\end{align}
\begin{align}
  \phi_0 \sigma_\phi/(1-\rho^2) =&   4\pi\mu^2 \exp \left\{ - {32\pi^2\left( 1-\rho^2 \right)^2 \over 3(N_c^2-1)\sigma_\phi}   -{1 \over 2}\right\} 
% \nonumber\\
%  =&   4\pi\mu^2 \exp \left\{ - {N_c(1-1/\sigma^2)^2 \over 12(N_c^2-1)(\sigma^2-1)}   {32\pi^2 \over g^2}  -{1 \over 2}\right\} 
  \nonumber\\
  =&    4\pi\mu^2 \exp \left\{ - {N_c^2 \over (N_c^2-1)}{(\sigma^2-1) \over \sigma^4}   {32\pi^2 \over 9\hat{g}^2}  -{1 \over 2}\right\} .
\label{phi0}
\end{align}
Note that the right-hand-side has a finite value in the limit $N_c \rightarrow \infty$. 
It turns out that $\sigma^2$ must be larger than one, $\sigma^2>1$.  
This result can be compared with the perturbative calculation of the effective potential to the three-loop order \cite{VKAV01}.
%%%%%%%%%%%%%%%%%%%%%%%%%%%%%%%%%%%%%%%%%%%%%%%%%%%%%%%%%%%%

%%%%%%%%%%%%%%%%%%%%%%%%%%%%%%%%%%%%%%%%%%%%%%%%%%%%%%%%%%%%

Now we estimate the order of the gluon mass by combining the above results with the lattice data. 
 Using the beta function at one-loop level in the relationship of the trace anomaly 
\begin{align}
  \langle 0| T_\mu{}^\mu |0 \rangle 
=& {\beta(\alpha_s) \over 4\alpha_s} \langle 0| (\mathscr{F}_{\mu\nu}^A)^2 |0 \rangle
%= {\beta(g) \over 2g} \langle 0| (\mathscr{F}_{\mu\nu}^A)^2 |0 \rangle
= {\pi \beta(\alpha_s) \over 4\alpha_s^2} \langle 0|{\alpha_s \over \pi} (\mathscr{F}_{\mu\nu}^A)^2 |0 \rangle
\nonumber\\  
\cong& - {11N_c-2N_f \over 24} \langle 0|{\alpha_s \over \pi} (\mathscr{F}_{\mu\nu}^A)^2 |0 \rangle ,
\end{align} 
and the values of the gluon condensation of mass dimension 4 calculated on a lattice~\cite{Ilgenfritz00}
\begin{align}
  \langle 0|{\alpha_s \over \pi} (\mathscr{F}_{\mu\nu}^A)^2 |0 \rangle= 0.152 \text{(GeV)}^4 (N_c=2), \quad
  0.144 \text{(GeV)}^4 (N_c=3)
\end{align}
we obtain
\begin{align}
   \langle 0| T_\mu{}^\mu |0 \rangle
= -0.1392 \text{(GeV)}^4 (N_c=2), \quad 
-0.1980 \text{(GeV)}^4 (N_c=3)
\end{align}
where we have put the flavor number $N_f=0$.

The effective potential just obtained has the minimum at $\phi=\phi_0$:
\begin{align}
  V(\phi_0) 
  = -{3(N_c^2-1) \over 128\pi^2} \left({\sigma_\phi \phi_0 \over 1-\rho^2}\right)^2
 = -{3(N_c^2-1) \over 128\pi^2}  m_{\mathscr{A}}^4 < 0 .
\end{align}
Supposing all the contribution to the vacuum energy comes only from the condensate $\phi_0$, i.e., 
\begin{align}
 V(\phi_0)  \cong {1 \over 4}  \langle 0| T_\mu{}^\mu |0 \rangle
 = -0.0495 \text{(GeV)}^4 (\text{for}~N_c=3) ,
\end{align}
 we obtain the rough estimate of the effective gluon mass 
\begin{equation}
 m_{\mathscr{A}}=\sqrt{\sigma_\phi \phi_0/(1-\sigma^{-2})}= 1.27 \ \text{GeV} .
\end{equation}
This value does not depend on the specific choice of the parameters $\rho$ and $\sigma$.  
The estimated value of  the effective gluon mass is larger than the result:   $m_{\mathscr{A}}=0.6~\text{GeV}$ of numerical simulation on a lattice \cite{LRG02}.
The vacuum condensate of mass dimension 2 is given by
\begin{align}
  | \langle g^2 \mathscr{A}_\mu^2 \rangle | =&   2 g^2 \phi_0 
=  2g^2 {1-\rho^2 \over \sigma_\phi} m_{\mathscr{A}}^2 
=   {N_c \over 2\sigma^2} m_{\mathscr{A}}^2 .
\end{align}

%In the large $N_c$ limit,
%\begin{align}
%  \phi_0 \sigma_\phi =   (1-\rho^2)4\pi\mu^2 \exp \left\{ -  {(\sigma^2-1) 32\pi^2 \over 12 \sigma^4}      -{1 \over 2}\right\} 
%\end{align}

%%%%%%%%%%%%%%%%%%%%%%%%%%%%%%%%%%%%%%%%%%%%%%%%%%%%%%%%%%%%%%%%%%%%%
%%%%%%%%%%%%%%%%%%%%%%%%%%%%%%%%%%%%%%%%%%%%%%%%%%%%%%%%%%%%%%%%%%%%%
%%%%%%%%%%%%%%%%%%%%%%%%%%%%%%%%%%%%%%%%%%%%%%%%%%%%%%%%%%%%%%%%%%%%%
%%%%%%%%%%%%%%%%%%%%%%%%%%%%%%%%%%%%%%%%%%%%%%%%%%%%%%%%%%%%%%%%%%%%%
%%%%%%%%%%%%%%%%%%%%%%%%%%%%%%%%%%%%%%%%%%%%%%%%%%%%%%%%%%%%%%%%%%%%%

\section{Static potential}

We introduce the source $J_{\mu\nu}$ of $B_{\mu\nu}$  by modifying the identity (\ref{id-B}) into 
\begin{align}
  1 = \int \mathcal{D}B_{\mu\nu} \exp \Big[ - \int d^4x 
{1 \over 4}  \{ B_{\mu\nu}-i {}^*[\rho (\partial_\mu \mathscr{A}_\nu - \partial_\nu \mathscr{A}_\mu) + \sigma g (\mathscr{A}_\mu \times \mathscr{A}_\nu)] + J_{\mu\nu}  \}^2 
\Big] .
\end{align}
This leads to the source term 
\begin{align}
   {1 \over 2}  B_{\mu\nu} J^{\mu\nu} 
-    {i \over 2} {}^*J^{\mu\nu} \rho (\partial_\mu \mathscr{A}_\nu - \partial_\nu \mathscr{A}_\mu) 
+  {1 \over 4}(J_{\mu\nu})^2 ,
\label{B-source}
\end{align} 
and the modification of $\mathscr{K}^{AB}_{\mu\nu}$ into 
$
  \mathscr{K}^{AB}_{\mu\nu} \rightarrow  \mathscr{K}^{AB}_{\mu\nu} - i \sigma g f^{ABC} {}^*J^{\mu\nu}{}^C 
$.

In the similar way, we can introduce the source $J^\phi, J^G{}_{\mu\nu}, J^\varphi{}^A, J^V{}_{\mu\nu}^A$ for the auxiliary field $\phi, G_{\mu\nu}, \varphi^A, V_{\mu\nu}^A$. Then the source term is further modified by adding 
\begin{align}
  & {1 \over 2}J^\phi \phi  + {1 \over 4} J^\phi \mathscr{A}_\mu \cdot \mathscr{A}^\mu + {1 \over 2\sigma_\phi} (J^\phi)^2 
\nonumber\\ & 
 + {1 \over 2}J^\varphi \cdot \varphi   + {1 \over 4} J^\varphi \cdot \mathscr{A}_\mu \star \mathscr{A}^\mu + {1 \over 2\sigma_\varphi} (J^\varphi)^2
\nonumber\\ & 
 + {1 \over 2}J_{\mu\nu}^G G^{\mu\nu} + {1 \over 4} J_{\mu\nu}^G S^{\mu\nu} + {1 \over 2\sigma_G} (J_{\mu\nu}^G)^2
\nonumber\\ & 
 + {1 \over 2} J_{\mu\nu}^V \cdot V^{\mu\nu}  + {1 \over 4} J_{\mu\nu}^V \cdot T^{\mu\nu} + {1 \over 2\sigma_V} (J_{\mu\nu}^V)^2  .
\end{align}
Thus, the introduction of the source terms for the auxiliary fields is equivalent to the replacement 
\begin{align}
  \mathscr{K}^{AB}_{\mu\nu} \rightarrow & \mathscr{K}^{AB}_{\mu\nu} - i \sigma g f^{ABC} {}^*J^{\mu\nu}{}^C 
 + {1 \over 2}\delta_{\mu\nu}\delta^{AB} J^\phi + {1 \over 2}\delta_{\mu\nu}d^{ABC} J^\varphi{}^C 
\nonumber\\&
+ {1 \over 2} \delta^{AB} \left(J^G_{\mu\nu} - {1 \over 4} \delta_{\mu\nu} J^G{}^\rho{}_\rho \right)  
+ {1 \over 2} d^{ABC} \left(J^G_C{}_{\mu\nu} - {1 \over 4} \delta_{\mu\nu} J^G_C{}^\rho{}_\rho \right) ,
\end{align}
and the addition of the terms
\begin{align}
  &   (\ref{B-source}) 
 + {1 \over 2}J^\phi \phi   
 + {1 \over 2}J^\varphi \cdot \varphi   
 + {1 \over 2}J_{\mu\nu}^G G^{\mu\nu}
 + {1 \over 2} J_{\mu\nu}^V \cdot V^{\mu\nu}  
\nonumber\\ & 
%+ {1 \over 4}(J_{\mu\nu})^2  
 + {1 \over 2\sigma_\phi} (J^\phi)^2  
+ {1 \over 2\sigma_\varphi} (J^\varphi)^2
+ {1 \over 2\sigma_G} (J_{\mu\nu}^G)^2 
 + {1 \over 2\sigma_V} (J_{\mu\nu}^V)^2  .
\end{align}

The linear term in the gluon field $\mathscr{A}_\mu$ appears only in the source term (\ref{B-source}) of $B_{\mu\nu}$ which is recast into the form 
\begin{align}
 \int d^4x  {1 \over 2}  B_{\mu\nu} \cdot J^{\mu\nu} 
- \int d^4x  \mathscr{A}_\mu \cdot J^\mu ,
\quad  J_\mu = -i \rho   \partial^\nu {}^*J_{\mu\nu} .
\label{cur}
\end{align}
which cancels the original source term 
$\int d^4x  \mathscr{A}_\mu \cdot J^\mu $ 
for $\mathscr{A}_\mu$. 
The current $J_\mu$  defined by (\ref{cur}) satisfies the conservation law
\begin{equation}
  \partial^\mu J_\mu = 0 .
\label{conserv}
\end{equation}
due to antisymmetry of the tensor current, $J_{\mu\nu}=-J_{\nu\mu}$.
  With the conservation law (\ref{conserv}), the $J_{\mu\nu}$ can be solved in terms of $J_\mu$:
\begin{align}
  J_{\mu\nu} = {1 \over i\rho} {1 \over n \cdot \partial} {}^*(n_\mu J_\nu - n_\nu J_\mu ) ,
  \label{rel}
\end{align}
where $n^\mu$ is a constant vector corresponding to a set of integration constants. 
Therefore, we can calculate the generating functional for the  source $J^\mu$ for the original gluon field $\mathscr{A}_\mu$ by making instead use of the antisymmetric tensor current source $J^{\mu\nu}$ coupled to the antisymmetric tensor field $B_{\mu\nu}$ via the relationship between two currents (\ref{rel}).

The integration over the field $B_{\mu\nu}$ is performed as follows. 
\begin{align}
 & \int \mathcal{D}B_{\mu\nu} \exp \left\{ - \int d^4x \left[  {1 \over 2} {}^*B^{\mu\nu}{}^A  
(\mathscr{D}^{AB}_{B_{\mu\nu},B_{\alpha\beta}})^{-1} 
 {}^*B^{\alpha\beta}{}^B  
+  {1 \over 2} {}^*B_{\mu\nu}^A {}^*J^{\mu\nu}{}^A \right] \right\} 
\nonumber\\&  
= \exp \left\{ - \int d^4x  {-1 \over 8} J^{\mu\nu}{}^A 
\mathscr{D}^{AB}_{B_{\mu\nu},B_{\alpha\beta}}
J^{\alpha\beta}{}^B   \right\} 
\nonumber\\&  
= \exp \left\{ - \int d^4x  {1 \over 8\rho^2}
\mathscr{D}_{B,B} 
{2n^2 \over (n \cdot \partial)^2}
J_\mu^A \left( \delta^{\mu\nu}-{n^\mu n^\nu \over n^2}  \right) J_\nu^A  \right\} 
\nonumber\\&  
= \exp \left\{ - \int {d^4k \over (2\pi)^4}  {-1 \over 4\rho^2}
\mathscr{D}_{B,B}(k)
{n^2 \over (n \cdot k)^2}
J_\mu^A(k) \left( \delta^{\mu\nu}-{n^\mu n^\nu \over n^2}  \right) J_\nu^A(-k) \right\} ,
\end{align}
where
$
 \mathscr{D}^{AB}_{B_{\mu\nu},B_{\alpha\beta}} 
= \mathscr{D}_{B,B} \delta^{AB} I_{\mu\nu,\alpha\beta} 
$.

%%%%%%%%%%%%%%%%%%%%%%%%%%%%%%%%%%%%%%%%%%%%%%%%%%%%%%%%%%%%

\begin{figure}[htbp]
\begin{center}
\includegraphics[height=5cm]{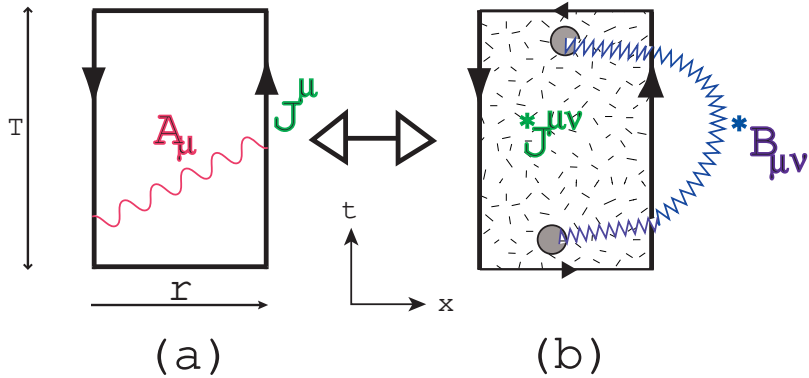}
\caption{Two equivalent descriptions for calculating the static potential.
(a) $J^\mu$ coupled to $\mathscr{A}_\mu$,
 (b) $J^{\mu\nu}$ coupled to $B_{\mu\nu}$. 
}
\label{fig:Coupling2suface}
\end{center}
\end{figure}

%%%%%%%%%%%%%%%%%%%%%%%%%%%%%%%%%%%%%%%%%%%%%%%%%%%%%%%%%%%%

We consider a pair of   opposite color charges $Q^A$ and $-Q^A$ located at $\bm{x}=\bm{a}$ and $\bm{x}=\bm{b}$, respectively.  In this case, the conserved current is given by 
$
 J_\mu^A(x) = Q^A \delta_{\mu 0}[\delta^3(\bm{x}-\bm{a})-\delta^3(\bm{x}-\bm{b})]
$. 
Using the Fourier transform  
\begin{align}
 J_\mu^A(k) = Q^A \delta_{\mu 0}2\pi \delta(k^0)
(e^{-i \bm{k}\cdot \bm{a}}-e^{-i \bm{k}\cdot \bm{b}}) ,
\end{align}
we obtain
\begin{align}
 & \int d^4x  {-1 \over 8} J^{\mu\nu}{}^A (x)
\mathscr{D}^{AB}_{B_{\mu\nu},B_{\alpha\beta}}
J^{\alpha\beta}{}^B(x)  
\nonumber\\&  
=  {1 \over 4\rho^2} Q^A Q^A \int dt \int {d^3k \over (2\pi)^3} 
\mathscr{D}_{B,B}(0,\bm{k})
{1 \over (\bm{n} \cdot \bm{k})^2}
(1-e^{i \bm{k}\cdot \bm{r}})(1-e^{-i \bm{k}\cdot \bm{r}}) ,
\label{JBJ}
\end{align}
where $\bm{r}:=\bm{a}-\bm{b}$ is the relative vector between a pair of charges, and we have chosen 
$n^\mu = (0, \bm{n})$ with a unit vector $\bm{n}$: $|\bm{n}|=1$.
Therefore the static potential is calculated from the formula: 
\begin{align}
 V_L(r) =  {\sigma^2 \over 2} Q^A Q^A \int {d^3k \over (2\pi)^3} 
\mathscr{D}_{B,B}(0,\bm{k})
{1 \over (\bm{n} \cdot \bm{k})^2}
 (1-e^{-i \bm{k}\cdot \bm{r}}) ,
\label{V_L}
\end{align}
where we have used 
$
 \mathscr{D}_{B,B}(0,-\bm{k})=\mathscr{D}_{B,B}(0,\bm{k})
$.

Substituting for 
$\mathscr{D}_{B,B}(0,\bm{k})$ 
the $k^0=0$ limit of the IR propagator (\ref{B-ir}): 
\begin{align}
  \mathscr{D}_{B,B}(k^2) 
=&    
%16\pi^2 (1-\sigma^{-2})^2 \sigma^{-2} f_1^{-1} 
 \left[ {1 \over 2(\sigma^2-1)}- {\hat{g}^2 \over 16\pi^2}{\sigma^2 \over (1-\sigma^{-2})^2} f_1 \right]^{-1} 
M^2 \left\{ m^2 + k^2 + O(k^4)   \right\}^{-1} ,
\nonumber\\
m^2 :=& M^2 \left[ {1 \over 2(\sigma^2-1)}- {\hat{g}^2 \over 16\pi^2}{\sigma^2 \over (1-\sigma^{-2})^2} f_1 \right]^{-1} 
  \left[ \frac{1}{2}-{\hat{g}^2 \over 16\pi^2}{\sigma^2 \over (1-\sigma^{-2})^2} f_0 \right]  , 
\nonumber\\
 M^2 :=& \phi_0/(1-\sigma^{-2}) ,
\label{B-ir2}
\end{align}
into the above equation (\ref{V_L}), we obtain 
the linear static potential as a function of $r:=|\bm{r}|$ for $r \gg 1$%
\footnote{
The vector $\bm{n}$ is chosen to be parallel to $\bm{r}$ by the axial symmetry of the system in question and the minimum energy condition. 
The method of calculations in what follows is the same as those of \cite{Suzuki88,SST95}. See there references for the details of calculations. 
} 
\begin{align}
 V_L(r) 
%=  {\sigma^2 \over 2} Q^A Q^A \int {d^3k \over (2\pi)^3} \mathscr{D}_{B,B}(0,\bm{k}) {1 \over (\bm{n} \cdot \bm{k})^2} (1-e^{-i \bm{k}\cdot \bm{r}})
= \sigma_{st} r ,
\end{align}
where the string tension $\sigma_{st}$ (with an enhance factor) is given by 
%\marginpar{RG inv.}
\begin{align}
  \sigma_{st} =&
%{8\pi^2 \over \hat{g}^2} (1-\sigma^{-2}) f_1^{-1}
    \sigma^2 \left[ {1 \over (\sigma^2-1)}- {\hat{g}^2 \over 8\pi^2}{\sigma^2 \over (1-\sigma^{-2})^2} f_1 \right]^{-1} 
{Q^A Q^A \over 8\pi} M^2 \ln \left( 1+{\Lambda^2 \over m^2} \right) 
\nonumber\\
=&     \sigma^4 \left[ 1 - {\hat{g}^2 \over 8\pi^2}{\sigma^6 \over (\sigma^{2}-1)} f_1 \right]^{-1} 
{Q^A Q^A \over 8\pi} \phi_0 \ln \left( 1+{\Lambda^2 \over m^2} \right) 
 ,
\end{align}
with
$\Lambda$ being the ultraviolet cutoff.%
\footnote{
This UV divergence originates from the fact that the VEV $\phi_0(x)=\langle \phi(x)  \rangle$ is a $x$-independent constant. 
In other words, the world sheet of the effective string connecting two color charges has no width. 
The presence of the UV cutoff $\Lambda$ means that the string has a non-zero width of the order  $\Lambda^{-1}$.  
See the next section.
} 
In the classical limit $\sigma \downarrow 1$, the string tension vanishes $\sigma_{st} \downarrow 0$. 
The string tension is essentially proportional to the vacuum condensate $\phi_0$. 
Thus, the non-vanishing string tension is generated from quantum effects. 
%\marginpar{Casimir scaling}

This string tension has the very similar form to that obtained by making use of Abelian projection assuming the Abelian dominance \cite{Suzuki88,SST95} up to a numerical factor, as expected from the type-II dual superconductor.%
\footnote{
It should be remarked that we have used the Abelian-like expression for the current $J_\mu$ for defining the Wilson loop.  
To incorporate the full non-Abelian structure, we must use the non-Abelian Stokes theorem which enables us to rewrite the line integral in the non-Abelian Wilson loop into the surface integral over the surface whose boundary coincides with the Wilson loop. Such a calculation was performed in the previous paper \cite{KI02}.
In this case, only the maximal torus subgroup is responsible for the final result, even if we start from the Lorentz gauge.  
Thus we reproduce the dual superconductor picture. 
}
The constant vector $n^\mu$ was introduced following the Zwanziger formulation \cite{Zwanziger70} which allows the simultaneous existence of electric and magnetic charge. 
Under the Abelian projection, only the Abelian charges corresponding to the maximal torus group contribute to the string tension. 
 In the Lorentz gauge, however, all the components of color charges $Q^A$ contribute to the string tension. 
Thus the propagation of $B_{\mu\nu}$ ($B_{0i}$ in this case) gives rise to the linear potential. 
Incidentally, in order to know the potential $V_L(r)$ for small $r$, we must use the UV asymptotic propagator of the form (\ref{B-uv}).

On the other hand, the original source term for the gluon field $\mathscr{A}_\mu$ 
\begin{align}
& \int d^4x  {-1 \over 2}J_\mu^A [\mathscr{K}_{\mu\nu}[\phi_0]\delta^{AB}]^{-1} J_\nu^B 
=  \int d^4x {-1 \over 2}J_\mu^A [-(1-\rho^2)\partial^2+\phi_0]^{-1}  J_\mu^A ,
\nonumber\\&  
=  - Q^A Q^A \int dt \int {d^3k \over (2\pi)^3} 
{1 \over (1-\rho^2)\bm{k}^2+\phi_0}
{1 \over 2}(1-e^{i \bm{k}\cdot \bm{r}})(1-e^{-i \bm{k}\cdot \bm{r}}) ,
\label{c}
\end{align}
yields the  Yukawa-type potential (with an enhance factor due to $\sigma$):
\begin{align}
  V_Y(r) = -(1-\sigma^{-2})^{-1} {Q^A Q^A \over 4\pi} {e^{-Mr} \over r} ,
\end{align}
apart from an irrelevant constant. 
The Yukawa-type potential does not depend on $n^\mu$ and it is gauge-parameter independent due to the conservation law (\ref{conserv}).

Here we comment the electric-magnetic dual picture of the Yang-Mills vacuum. 
The dual transformation (or Fourier transformation) can be used to transform the massive antisymmetric tensor field theory into another equivalent theory.  
It is well known \cite{Kawai81} that in $D$-dimensional space-time, a {\it massless  antisymmetric tensor field} of rank $p$ ($p$-form) is dual to the $(D-p-2)$-form, while a {\it massive  antisymmetric tensor field} of rank $p$ is dual to the $(D-p-1)$-form.  
In fact, it is shown that the effective theory  retained up to the quadratic terms in the massive Kalb-Ramond \cite{KR74} field $B_{\mu\nu}$ is dual to the massive vector field $U_\mu$ theory described by the dual Ginzburg-Landau (DGL) theory in the London limit. 
See \cite{Kondo00,KI02,DK02} for details.
Therefore, the quite similar situation to the dual superconductivity is realized in the Yang-Mills vacuum  in the $1/N_c$ expansion and the the London limit corresponds to in the leading order of the expansion. 
  To go beyond the London limit, it it necessary \cite{Kondo00,KI02} to incorporate the quartic term in $B_{\mu\nu}$ which appears in the next-to-leading order of the $1/N_c$ expansion.

%%%%%%%%%%%%%%%%%%%%%%%%%%%%%%%%%%%%%%%%%%%%%%%%%%%%%%%%%%%%%%%%%%%%%
%%%%%%%%%%%%%%%%%%%%%%%%%%%%%%%%%%%%%%%%%%%%%%%%%%%%%%%%%%%%%%%%%%%%%
%%%%%%%%%%%%%%%%%%%%%%%%%%%%%%%%%%%%%%%%%%%%%%%%%%%%%%%%%%%%%%%%%%%%%
%%%%%%%%%%%%%%%%%%%%%%%%%%%%%%%%%%%%%%%%%%%%%%%%%%%%%%%%%%%%%%%%%%%%%
%%%%%%%%%%%%%%%%%%%%%%%%%%%%%%%%%%%%%%%%%%%%%%%%%%%%%%%%%%%%%%%%%%%%%
\section{String representation}

In the derivation of static potential in the previous section, we adopted a specific configuration for the color charges and introduced a constant vector $n^\mu$.  These shortcoming
 are avoided and we can derive more general result by means of the string representation of Yang-Mills theory as will be shown in what follows. 

The trajectory $C$ of the charge $Q^A$ is parameterized by a parameter $\tau$. Hence the current $J^\mu$ is written as
\begin{align}
 J_\mu^A(x) = Q^A \int_{C} dy_\mu(\tau) \delta^4(x-y(\tau)) 
= \int_{C} d\tau {dy_\mu(\tau) \over d\tau} \delta^4(x-y(\tau)) .
\end{align}
Let the trajectory be a loop $C$ and let $\Sigma$ be the surface whose boundary is $C$, i.e., $\partial \Sigma=C$.
The surface $\Sigma$ is parameterized by two parameters $\tau, \sigma$.  
Then it is shown that the vorticity tensor current defined by 
\begin{align}
  \Theta_{\mu\nu}^A(x) :=& Q^A \int \int_{\Sigma} d^2 S_{\mu\nu}(y(\tau,\sigma)) \delta^4(x-y(\tau,\sigma)) 
\nonumber\\
=& \int \int_{\Sigma} d\tau d\sigma {\partial(y_\mu,y_\nu) \over \partial(\tau,\sigma)} \delta^4(x-y(\tau,\sigma)) ,
\end{align}
satisfies the relationship 
\begin{align}
  \partial^\nu \Theta_{\mu\nu}^A(x) = J_\mu^A(x) .
\end{align}
The $\Theta_{\mu\nu}^A(x)$ has its support on the two-dimensional surface $\Sigma$, while the $J_\mu^A(x)$ has the support on the loop $C$ in the four-dimensional target space $x_\mu$. 
As in the previous section, we have the correspondence
\begin{align}
  J_{\mu\nu}^A(x) = i \rho^{-1}\ {}^*\Theta_{\mu\nu}^A(x)  .
\end{align}
The contribution of the vorticity current is estimated as 
\begin{align}
 &  \int \mathcal{D}B_{\mu\nu}  \exp \left\{ - \int d^4x \left[  {1 \over 2} {}^*B^{\mu\nu}{}^A  
(\mathscr{D}^{AB}_{B_{\mu\nu},B_{\alpha\beta}})^{-1} 
 {}^*B^{\alpha\beta}{}^B  
+  {1 \over 2} B_{\mu\nu}^A J^{\mu\nu}{}^A \right] \right\} 
\nonumber\\&  
= \int \mathcal{D}B_{\mu\nu} \exp \left\{ - \int d^4x   \left[  {1 \over 2} B^{\mu\nu}{}^A  
(\mathscr{D}^{AB}_{B_{\mu\nu},B_{\alpha\beta}})^{-1} 
 B^{\alpha\beta}{}^B  
+  {i \rho^{-1} \over 2} B_{\mu\nu}\ {}^*\Theta_{\mu\nu}^A \right] \right\}
\nonumber\\&  
= \exp \left\{ - \int d^4x \int d^4x'  {1 \over 8\rho^2}
\Theta_{\mu\nu}^A(x) 
\mathscr{D}^{AB}_{B_{\mu\nu},B_{\alpha\beta}}(x,x')
 \Theta_{\alpha\beta}^B(x')  \right\}  
\nonumber\\&  
= \exp \left\{ -  {Q^2 \over 8\rho^2}
\int_{\Sigma} d^2S_{\mu\nu}(x(\sigma)) \int_{\Sigma} d^2S_{\alpha\beta}(x'(\sigma')) \mathscr{D}_{B_{\mu\nu},B_{\alpha\beta}}
(x(\sigma),x'(\sigma')) 
  \right\} .
\label{NLaction}
\end{align}

The action in (\ref{NLaction}) describes the interaction between two surface-elements $dS_{\mu\nu}$ and $dS_{\alpha\beta}$ in the surface $\Sigma$ spanned by the loop $C$.
This expression (\ref{NLaction}) is the same as the non-local action for the vorticity tensor current, from which the string representation of the Wilson loop average in Yang-Mills theory was derived in \cite{Kondo00,KI02} by using the method \cite{Antonov99}.  We can show 
\begin{subequations}
\begin{align}
 (\ref{NLaction}) =&  \exp (-S_{cs}[x]) ,
  \nonumber\\
    S_{cs} =& \sigma_{st} \int_{\Sigma} d^2\sigma \sqrt{g} 
  + \alpha_0^{-1} \int_{\Sigma} d^2\sigma \sqrt{g}  g^{ab} \partial_a t_{\mu\nu} \partial_b t_{\mu\nu} + \kappa_t \int_{\Sigma} d^2\sigma \sqrt{g}R + \cdots ,
\label{Stringaction}
\end{align}
where
\begin{align}
 \sigma_{st}  =&  \sigma^4 \left[ 1 - {\hat{g}^2 \over 8\pi^2}{\sigma^6 \over (\sigma^{2}-1)} f_1 \right]^{-1} 
{Q^A Q^A \over 8\pi} \phi_0 \ln \left( 1+{\Lambda^2 \over m^2} \right),
 \\
 \alpha_0^{-1} =& -  {\pi \over 2} Q^2  (1-\sigma^{-2})^2  <0  ,
 \\
 \kappa_t =& {\pi \over 3} Q^2  (1-\sigma^{-2})^2   > 0 .
\end{align}
\end{subequations}
In the action (\ref{Stringaction}), the first term is the Nambu-Goto action, the second term is the rigidity term, and 
the third term is the intrinsic curvature term.
See  \cite{Kondo00,KI02} for the details of the calculation. 
The string theory just derived is nothing but the rigid string with a negative rigidity term proposed by Polyakov \cite{Polyakov96}.
Thus we have shown that the vacuum condensate of mass dimension two gives a contribution to 
the non-vanishing string tension $\sigma_{st}$ in the linear potential already in the leading order of $1/N_c$ expansion.

%%%%%%%%%%%%%%%%%%%%%%%%%%%%%%%%%%%%%%%%%%%%%%%%%%%%%%%%%%%%%%%%%%%%%
%%%%%%%%%%%%%%%%%%%%%%%%%%%%%%%%%%%%%%%%%%%%%%%%%%%%%%%%%%%%%%%%%%%%%
%%%%%%%%%%%%%%%%%%%%%%%%%%%%%%%%%%%%%%%%%%%%%%%%%%%%%%%%%%%%%%%%%%%%%
%%%%%%%%%%%%%%%%%%%%%%%%%%%%%%%%%%%%%%%%%%%%%%%%%%%%%%%%%%%%%%%%%%%%%
%%%%%%%%%%%%%%%%%%%%%%%%%%%%%%%%%%%%%%%%%%%%%%%%%%%%%%%%%%%%%%%%%%%%%

\section{Conclusion and discussion}

We have proposed an equivalent formulation of the Yang-Mills theory in which all the gluon self-interactions are absent. 
In other words, the gluon fields are at most quadratic in the new formulation and all the gluon fields can be integrated out exactly in the $SU(N_c)$ Yang-Mills theory to obtain an equivalent formulation at the cost of  introducing various auxiliary fields, i.e., an antisymmetric tensor $B_{\mu\nu}^A$, two symmetric tensors $G_{\mu\nu}, V_{\mu\nu}^A$, four Lorentz scalars $\phi, \varphi^A, \phi_{FP}, \varphi_{FP}^A$. 
Hence the resulting theory is written in terms of the seven auxiliary fields and the ghost $\mathscr{C}$ and antighost fields $\bar{\mathscr{C}}$. 
Two of the auxiliary fields $\phi_{FP}, \varphi_{FP}^A$ are necessary only when the four-ghost interaction exists.

Then we have  set up a new $1/N_c$ expansion on a non-trivial vacuum in which we have shown that one of the scalar field $\phi$  can acquire a non-vanishing vacuum expectation value
$\langle \phi \rangle \equiv \phi_0 \not=0$.
  This corresponds to the on-shell BRST-invariant vacuum condensate of mass dimension two \cite{Kondo01,KMSI02,Kondo03b} which reduces in the Landau gauge $\lambda=0$ 
to the gluon pair condensation 
$\langle \mathscr{A}_\mu^2 \rangle\not=0$. 
Once such type of Bose-Einstein condensation occurs,  all the gluons acquire a common effective mass $M$. 
This behavior of the gluon propagator is consistent with the recent results of the Schwinger-Dyson equation \cite{SDE03}  and the axiomatic approach \cite{Kondo03a}.

In the condensed vacuum, all the auxiliary fields acquire their kinetic term and consequently become dynamical fields.
We have given the  Feynman rules of the $1/N_c$ expansion which give the propagators of gluon fields, auxiliary fields and (anti)ghost fields, and the interaction vertices among the gluon and other fields.  
The resulting theory is expected to be renormalizable in the scheme of the $1/N_c$ expansion just obtained.  However, the detailed analyses of the renormalizability  will be postponed to a subsequent paper. 

The effective gluon mass has been estimated in the leading order by assuming that the vacuum energy is saturated by the vacuum condensate $\phi_0$ alone. 
The obtained value is larger than the expected value due to numerical simulations on a lattice. 
This might suggest that we must go beyond the leading order or take into account other vacuum condensates. 
However, the result does not contradict with the experimental upper bound \cite{Field02}.
Moreover, the renormalization of the composite operator $\mathscr{A}_\mu^2$ in the given $1/N_c$ expansion is to be investigated in future works. 

As a by-product, the mass  of the lightest scalar glueball mass to the gluon mass is obtained by performing the partial summation of the leading order diagrams of the $1/N_c$ expansion:
$M(0^{++})/M=\sqrt{6} \cong 2.45$ independently of the parameter $\sigma$. This value agrees with that of the potential model by Cornwall and Soni \cite{CS83,HLW01}. 
In order to reproduce the expected mass of the glueball $0^{++}$, i.e., $M(0^{++}) = 1.5 \sim 1.7 \text{GeV}$,  the gluon must acquire the mass $M = 600 \sim 700 \text{MeV}$ in agreement with the numerical simulation on a lattice \cite{LRG02}. 
It is desirable to calculate the complete glueball spectrum in the future work.

The static quark potential calculated in the same approximation  is expressed as a combination of the Yukawa type potential and the linear potential. 
The non-vanishing string tension is provided by the vacuum condensate of mass dimension two proposed by \cite{Kondo01}. 
It is argued that the  mixed gluon-ghost condensate as the VEV of the on-shell BRST invariant composite operator of mass dimension two  \cite{Kondo01} means the {\it transverse} gluon pair condensation of mass dimension 2, which is gauge-invariant but nonlocal \cite{Kondo03b}. 
In the Maximal Abelian gauge, these results are consistent with the dual superconductor picture of QCD vacuum examined through the string representation \cite{Kondo00,KI02}.

The obtained string tension $\sigma_{st}$ depends on the parameter $\sigma$, just as the gluon mass $M$ does. This result should be compared with the other large $N$ results, e.g., the $N=\infty$ extrapolation of the SU(2) and SU(3) simulation results on the lattice  \cite{Teper98}:
$M(0^{++})/\sqrt{\sigma_{st}} \cong 3.56\pm0.18, \cdots$,  
or the light-front Hamiltonian approach of the transverse lattice gauge theory \cite{DS98}:
$M(0^{++})/\sqrt{\sigma_{st}} \cong 3.3\pm0.2, \cdots$. 
In the phenomenological applications, the parameter $\sigma$ and $\Lambda$ (in the string tension) can be adjusted to reproduce the gluon mass $M = 600 \sim 700 \text{MeV}$ and the string tension $\sqrt{\sigma_{st}}=450 \text{MeV}$. 
By adopting these values as inputs, one can calculate the complete glueball spectrum and other physical quantities. 

It is also interesting to study the fate of the pair condensation in relation with deconfinement, if the sufficient number $N_f$ of quark flavors are incorporated into the theory.  It is interesting to confirm whether or not the pair condensation disappears above a critical number of quark flavors \cite{GM78}.
Some phenomenological consequences due to the vacuum condensate of mass dimension two were discussed also in the papers \cite{phenomeno}.

{\bf Note added}:
After having submitted this paper to e-Print archive, the author was informed  that some similar results were previously obtained  on the relation among the 
condensate of dimension 2, the gluon condensate and the mass of 
the dual gluon, which is related to the glueball \cite{BBDV98}
(but just inside two models, i.e., stochastic vacuum model \cite{SVM} and dual long-distance QCD \cite{dQCD}, and not 
with a complete theory derivation as this paper).
The author would like to thank Nora Brambilla for sending this information.

%\marginpar{Instanton vs. 1/N}
%\marginpar{$x$-dep. $\phi_0$}

%%%%%%%%%%%%%%%%%%%%%%%%%%%%%%%%%%%%%%%%%%%%%%%%%%%%%%%%%%%%%%%%%%%%%
%%%%%%%%%%%%%%%%%%%%%%%%%%%%%%%%%%%%%%%%%%%%%%%%%%%%%%%%%%%%%%%%%%%%%
%%%%%%%%%%%%%%%%%%%%%%%%%%%%%%%%%%%%%%%%%%%%%%%%%%%%%%%%%%%%%%%%%%%%%
%%%%%%%%%%%%%%%%%%%%%%%%%%%%%%%%%%%%%%%%%%%%%%%%%%%%%%%%%%%%%%%%%%%%%
%%%%%%%%%%%%%%%%%%%%%%%%%%%%%%%%%%%%%%%%%%%%%%%%%%%%%%%%%%%%%%%%%%%%%

\appendix
%\newpage
\section{Cancellation of divergence  and renormalization}

First, we focus on the divergence in 
\begin{align}
 {1 \over 3(\sigma^2-1)\hat{g}^2} \phi_0 
+{1 \over 2} {\rm tr} (\mathscr{K}^{-1}_{\mu\nu}[\phi_0])   .
\label{gap}
\end{align}
The dimensional regularization yields 
\begin{align}
  \int {d^D k \over (2\pi)^D} {1 \over [k^2 + \Phi_0]^\alpha} = {\Gamma(\alpha-{D \over 2}) \over (4\pi)^{D/2} \Gamma(\alpha)} (\Phi_0)^{-\alpha+{D \over 2}} 
  = {\Gamma(\epsilon-2+\alpha) \over (4\pi)^{2-\alpha-\epsilon}} (\Phi_0)^{2-\alpha-\epsilon} ,
\label{formula}
\end{align}
where $\epsilon:=2-D/2$. 
Defining 
$
 \Phi_0 := \phi_0/(1-\rho^2)
$
and 
$\tilde\lambda:=(1-\rho^2)\lambda$, 
we have
\begin{align}
 & (1-\rho^2) {\rm tr} (\mathscr{K}^{-1}_{\mu\nu}[\phi_0])
\nonumber\\ &
  =   \int {d^D k \over (2\pi)^D} \left[ {D-1 \over k^2 + \Phi_0} +  {\tilde\lambda \over k^2 + \tilde\lambda \Phi_0} \right] 
\nonumber\\ &
=   {\Gamma(\epsilon-1) \over (4\pi)^{2-\epsilon}} \Phi_0 [D-1+\tilde\lambda^{2-\epsilon} ] (\Phi_0)^{-\epsilon} 
%\nonumber\\ &
%=    {\Gamma(\epsilon-1) \over (4\pi)^{2}} \Phi_0 [D-1+\tilde\lambda^{2} e^{-\epsilon \ln \tilde\lambda } ] e^{\epsilon \ln (4\pi)-\epsilon \ln (\Phi_0)} 
%\nonumber\\ &
%=    {\Phi_0 \over (4\pi)^{2}}  [-\epsilon^{-1}+\gamma-1+O(\epsilon)] 
% [1+\epsilon \ln (4\pi)-\epsilon \ln (\Phi_0)+O(\epsilon^2) ]
%\nonumber\\ & \times
%[D-1+\tilde\lambda^{2} e^{-\epsilon \ln \tilde\lambda } ] 
%\nonumber\\ &
%=   {\Phi_0 \over (4\pi)^{2}}  [-\epsilon^{-1}+\gamma- \ln (4\pi)-1+\ln \Phi_0+O(\epsilon)][D-1+\tilde\lambda^{2} -\epsilon \tilde\lambda^{2} \ln \tilde\lambda + O(\epsilon^2)] 
\nonumber\\ &
=   {\Phi_0 \over (4\pi)^{2}}  [-\bar\epsilon^{-1}(D-1+\tilde\lambda^{2})+(D-1+\tilde\lambda^{2})(\ln \Phi_0-1)+\tilde\lambda^{2} \ln \tilde\lambda + O(\epsilon) ]  ,
\end{align}
where we have defined $\bar\epsilon^{-1}:=\epsilon^{-1}-\gamma+ \ln (4\pi)$ and used the Laurent expansion of the Gamma function
$
 \Gamma(\epsilon-1) = -\epsilon^{-1}+\gamma-1+O(\epsilon)
$.  
Then we obtain
\begin{align}
   {1 \over 2} {\rm tr} (\mathscr{K}^{-1}_{\mu\nu}[\phi_0])
 =& (1-\rho^2)^{-2} {\phi_0 \over 2(4\pi)^{2}}  \Big[-{D-1+\tilde\lambda^{2} \over \bar\epsilon} 
\nonumber\\ &
    + (D-1+\tilde\lambda^{2}) \left( \ln {\phi_0 \over 1-\rho^2}-1 \right)+\tilde\lambda^{2} \ln \tilde\lambda + O(\epsilon) \Big] .
\end{align}
Therefore, the logarithmic divergence of the second term should be absorbed into the first term, 
\begin{align}
 {\rm divergent~part~of~}{1 \over 3(\sigma^2-1)\hat{g}^2} \phi_0 
 - (1-\rho^2)^{-2} {\phi_0 \over 2(4\pi)^{2}}  {D-1+\tilde\lambda^{2} \over \bar\epsilon}  = 0 .
 \label{divcan}
\end{align}

Next, in order to calculate the divergent part in the second term of
\begin{align}
  {1 \over 3(\sigma^2-1)\hat{g}^2}  -{1 \over 2} \int {d^Dp \over (2\pi)^D}  \mathscr{K}^{-1}_{\mu\nu}[p;\phi_0]  \mathscr{K}^{-1}_{\nu\mu}[p+k;\phi_0] ,
\end{align}
 we put $k=0$.  
Then, using the integration formula (\ref{formula}), 
%\begin{align}
% \int {d^D k \over (2\pi)^D} {1 \over [k^2 + \Phi_0]^2} = {\Gamma(2-{D \over 2}) \over (4\pi)^{D/2} \Gamma(2)} (\Phi_0)^{-2+{D \over 2}} 
%  = {\Gamma(\epsilon) \over (4\pi)^{2-\epsilon}} (\Phi_0)^{-\epsilon} ,
%\end{align}
the second term reads 
\begin{align}
 &  (1-\rho^2)^{2}{\rm tr} (\mathscr{K}^{-1}_{\mu\nu}[\phi_0] \mathscr{K}^{-1}_{\mu\nu}[\phi_0])
\nonumber\\ &
  =  \int {d^D k \over (2\pi)^D} \left[ {D-1 \over (k^2 + \Phi_0)^2} +  {\tilde\lambda \over (k^2 + \tilde\lambda \Phi_0)^2} \right] 
\nonumber\\ &
=   {\Gamma(\epsilon) \over (4\pi)^{2-\epsilon}}   (\Phi_0)^{-\epsilon}  [D-1+\tilde\lambda^{2-\epsilon} ]
%\nonumber\\ &
%=   {\Gamma(\epsilon) \over (4\pi)^{2}}  e^{\epsilon \ln (4\pi)-\epsilon \ln (\Phi_0)} [D-1+\tilde\lambda^{2} e^{-\epsilon \ln \tilde\lambda } ] 
%\nonumber\\ &
%=    {1 \over (4\pi)^{2}}  [\epsilon^{-1}-\gamma+O(\epsilon)] 
% [1+\epsilon \ln (4\pi)-\epsilon \ln (\Phi_0)+O(\epsilon^2) ]
%\nonumber\\ & \times
%[D-1+\tilde\lambda^{2} e^{-\epsilon \ln \tilde\lambda } ] 
%\nonumber\\ &
%=    {1 \over (4\pi)^{2}}  [\epsilon^{-1}-\gamma+ \ln (4\pi)-\ln \Phi_0+O(\epsilon)][D-1+\tilde\lambda^{2} -\epsilon \tilde\lambda^{2} \ln \tilde\lambda + O(\epsilon^2)] 
\nonumber\\ &
=   {1 \over (4\pi)^{2}}  [\bar\epsilon^{-1}(D-1+\tilde\lambda^{2})-(D-1+\tilde\lambda^{2})\ln \Phi_0-\tilde\lambda^{2} \ln \tilde\lambda + O(\epsilon) ]  ,
\end{align}
where we have used 
$
 \Gamma(\epsilon) = \epsilon^{-1}-\gamma+O(\epsilon) 
$.
Therefore, the logarithmic divergence of the second term is cancelled with that of the first term ${1 \over 3(\sigma^2-1)}$, 
\begin{align}
 {\rm divergent~part~of~}{1 \over 3(\sigma^2-1)\hat{g}^2}  
 - (1-\sigma^{-2})^{-2} {1 \over 2(4\pi)^{2}}  {D-1+\tilde\lambda^{2} \over \bar\epsilon}  = 0 ,
\end{align}
provided that   the relation (\ref{divcan}) is satisfied. 
Therefore, the auxiliary field  propagator $\mathscr{D}_{\tilde\phi\tilde\phi}(k^2)$ has the well-defined finite UV expression.

In order to absorb the divergent part in the relation (\ref{divcan}), we perform the renormalization.  We introduce the renormalization factor $Z_\sigma$ by 
$\sigma=Z_\sigma \sigma_R$ to absorb the divergence in such a way that  
\begin{align}
 {\rm divergent~part~of~}{1 \over 3(Z_\sigma^2 \sigma_R^2-1)}  
 = {Z_\sigma^4 \sigma_R^4 \over (Z_\sigma^2 \sigma_R^2-1)^2} {\hat{g}^2 \over 2(4\pi)^{2}}  {D-1+\tilde\lambda^{2} \over \bar\epsilon}    .
\end{align}
 Taking into account $Z_\sigma=1+Z_\sigma^{(1)}+\cdots$ and 
\begin{align}
 {1 \over 3(Z_\sigma^2 \sigma_R^2-1)}  
=  {1 \over 3(\sigma_R^2-1)} -  Z_\sigma^{(1)}{2\sigma_R^2 \over 3(\sigma_R^2-1)^2} + \cdots  ,
\end{align}
the renormalization factor is determined as 
\begin{align}
Z_\sigma^{(1)}
=  -   \frac{3}{4}{\mu^{-2\epsilon}\hat{g}^2 \over (4\pi)^{2}}  \sigma_R^2 {D-1+\tilde\lambda^{2} \over \bar\epsilon} ,
\end{align}
where we have introduced the renormalization scale $\mu$ to render the coupling constant dimensionless for arbitrary $D$. The  RG function is calculated as 
\begin{align}
 \gamma_\sigma := {1 \over 2} \mu{\partial \sigma_R \over \partial \mu} = - {1 \over 2} \sigma_R {\partial \ln Z_\sigma \over \partial \ln \mu} \cong - {1 \over 2} \sigma_R {\partial Z_\sigma^{(1)} \over \partial \ln \mu} 
 = -\frac{3}{4}{\hat{g}^2 \over (4\pi)^{2}} \sigma_R^3   (D-1+\tilde\lambda^{2}) <0  . 
\end{align}
This implies that $\sigma_{R}$ increases as $\mu$ decreases. 
This tendency is favorable to our choice $\sigma_R^2>1$. 
The explicit solution $\sigma_{R}$ of this equation for $D=4$ and $\tilde\lambda=0$ is given as
\begin{align}
  \sigma_{R}(\mu)  = \sigma_{R}(\mu_0) \left[ 1+ \sigma_{R}(\mu_0){27 \over 22} \hat{g}^2 \ln {\ln \mu/\Lambda \over \ln \mu_0/\Lambda } \right]^{-1} 
  \cong  {22 \over 27} \hat{g}^{-2} \left[ \ln {\ln \mu/\Lambda \over \ln \mu_0/\Lambda } \right]^{-1} , 
\end{align}
by making use of the one-loop expression of the running coupling constant \cite{KMSI02}.
This equation is valid when $\Lambda \ll \mu < \mu_0$, i.e., for small  't~Hooft coupling $\hat{g}^2$. 
Note that the  $N_c$ dependence enters in this expression through the 't~Hooft coupling $\hat{g}^2$. 

At a certain value of $\mu$, say, $\mu_0$, we can choose an arbitrary value of $\sigma_R$, say, $\sigma_0:=\sigma_R(\mu_0)$.  Then the value of $\sigma_R(\mu)$ at another point $\mu$ is calculated according to the RG equation. Here the initial value $\rho_R(\mu_0)$ of $\rho_R(\mu)$ at $\mu=\mu_0$ must be tuned  so that $\rho_R(\mu)=\sigma_R^{-1}(\mu)$ at the desired value of $\mu$, since the gluon self-interactions are eliminated only when this relation is satisfied. 
In the $1/N_c$ expansion, the magnitude of 't~Hooft coupling $\hat{g}^2$ can not be specified, although it is expected to be small in our treatment.  

%%%%%%%%%%%%%%%%%%%%%%%%%%%%%%%%%%%%%%%%%%%%%%%%%%%%%%%%%%%%%%%%%%%%%
%%%%%%%%%%%%%%%%%%%%%%%%%%%%%%%%%%%%%%%%%%%%%%%%%%%%%%%%%%%%%%%%%%%%%
%%%%%%%%%%%%%%%%%%%%%%%%%%%%%%%%%%%%%%%%%%%%%%%%%%%%%%%%%%%%%%%%%%%%%
%%%%%%%%%%%%%%%%%%%%%%%%%%%%%%%%%%%%%%%%%%%%%%%%%%%%%%%%%%%%%%%%%%%%%
%%%%%%%%%%%%%%%%%%%%%%%%%%%%%%%%%%%%%%%%%%%%%%%%%%%%%%%%%%%%%%%%%%%%%

\section{Calculation of loop integrals}

The dimensional regularization in the modified minimal subtraction scheme ($\overline{\text{MS}}$) ($\bar\epsilon^{-1}:=2/(4-D)-\gamma_E+\ln 4\pi$) leads to  the following results in the Euclidean formulation.  In following equations, $k^2$ denotes the Euclidean momentum squared.  The corresponding expression in the Minkowski region is obtained by $k^2 \rightarrow -k^2$ below the threshold $k^2<4M^2$. 
It is possible to obtain the full expression for arbitrary $\lambda$. But it is quite complicated and long. Hence we write down two special but important cases explicitly. 

(1) In the Landau gauge $\lambda=0$:
\begin{align}
\Pi_{\mu\nu,\alpha\beta}^{AB}(k) 
 =&  
  \delta^{AB}  { g^2 N_c    \over 16\pi^2}  {\sigma^2 \over (1-\rho^2)^2}
\Biggr\{ I_{\mu\nu,\alpha\beta} \Biggr[  \epsilon^{-1} {1 \over 2} 
-{17 \over 18} + {1 \over 2} \ln {M^2 \over \mu^2} - {5 \over 18}{k^2 \over M^2} +{1 \over 2}{M^2 \over k^2} 
\nonumber\\ &
+{1 \over 6} {(k^2-2M^2) \over k^2} \sqrt{1+{4M^2 \over k^2}} \ln {\sqrt{1+{4M^2 \over k^2}}+1 \over \sqrt{1+{4M^2 \over k^2}}-1} 
\nonumber\\
 & +{1 \over 6}{(k^2+M^2)^3 \over k^4 M^2} \ln \left( 1+{k^2 \over M^2} \right) 
 \Biggr] 
+ {1 \over 2}(I-P)_{\mu\nu,\alpha\beta} \left[ ... \right]
\Biggr\}   ,
\end{align}
where  
$M^2:=\phi_0/(1-\rho^2)$
and we have introduced
\begin{equation}
 P_{\mu\nu,\alpha\beta} := {1 \over 2}(P^T_{\mu\alpha}P^T_{\nu\beta}-P^T_{\mu\beta}P^T_{\nu\alpha}), \quad 
P^T_{\mu\nu} := \delta_{\mu\nu} - {k_\mu k_\nu \over k^2} .
\end{equation}
In the large  momentum $k^2$ region, on the other hand, the propagator reads 
\begin{align}
\Pi_{\mu\nu,\alpha\beta}^{AB}(k) 
= \delta^{AB}  { g^2 N_c    \over 16\pi^2}  {\sigma^2 \over (1-\rho^2)^2}  I_{\mu\nu,\alpha\beta} &
\Big[  \epsilon^{-1} {1 \over 2} 
+  {k^2 \over M^2} \left( {1 \over 6}\ln {k^2 \over M^2} -{5 \over 18} \right) 
\nonumber\\&
 -{7 \over 9} +{1 \over 2}\ln {M^2 \over \mu^2}+{2 \over 3}\ln {k^2 \over M^2} + O(1/k^2) \Big] .
\end{align}
Here we have used the following properties of the projection operator $I_{\mu\nu,\alpha\beta}$:
\begin{subequations}
\begin{align}
I_{\mu\nu,\alpha\beta} =& - I_{\nu\mu,\alpha\beta} = I_{\alpha\beta,\mu\nu}
= - I_{\mu\nu,\beta\alpha} ,
\\
I_{\mu\nu,\rho\sigma}I_{\rho\sigma,\alpha\beta} =& I_{\mu\nu,\alpha\beta}, 
\\
k^{\rho_1} I_{\mu\nu,\rho_1\sigma} k^{\rho_2} I_{\alpha\beta,\rho_2\sigma}
=& {1 \over 4}(k_\mu k_\alpha \delta_{\nu\beta}-k_\mu k_\beta \delta_{\nu\alpha}
-k_\nu k_\alpha \delta_{\mu\beta}+k_\nu k_\beta \delta_{\mu\alpha})
\nonumber\\
=& {1 \over 2}k^2 (I-P)_{\mu\nu,\alpha\beta} ,
\\
 I_{\mu\nu,\alpha\beta}B^{\alpha\beta}= & B_{\mu\nu},
\quad
 B^{\mu\nu}I_{\mu\nu,\alpha\beta}=B_{\alpha\beta} ,
\end{align}
\end{subequations}
for an arbitrary antisymmetric tensor $B_{\mu\nu}$.

(2) In the Feynman gauge $\lambda=1$:
\begin{align}
\Pi_{\mu\nu,\alpha\beta}^{AB}(k) 
 =&  
  \delta^{AB}  { g^2 N_c    \over 16\pi^2}  {\sigma^2 \over (1-\rho^2)^2}
\Biggr\{ I_{\mu\nu,\alpha\beta} \Biggr[  \epsilon^{-1}  
-{10 \over 9} +   \ln {M^2 \over \mu^2} - {5 \over 18}{k^2 \over M^2} +{2 \over 3}{M^2 \over k^2} 
\nonumber\\ &
+{1 \over 12} {(2k^2-M^2) \over M^2} \left( 1+{4M^2 \over k^2} \right)^{3/2} \ln {\sqrt{1+{4M^2 \over k^2}}+1 \over \sqrt{1+{4M^2 \over k^2}}-1} 
  \Biggr] 
\nonumber\\
 &+ {1 \over 2}(I-P)_{\mu\nu,\alpha\beta} \left[ ... \right]
\Biggr\}   ,
\end{align}
In the large  momentum $k^2$ region, on the other hand, 
the propagator reads 
\begin{align}
\Pi_{\mu\nu,\alpha\beta}^{AB}(k) 
= \delta^{AB}  { g^2 N_c    \over 16\pi^2}  {\sigma^2 \over (1-\rho^2)^2}  I_{\mu\nu,\alpha\beta} &
\Big[  \epsilon^{-1}  
+  {k^2 \over M^2} \left( {1 \over 6}\ln {k^2 \over M^2} -{5 \over 18} \right) 
\nonumber\\&
 -{7 \over 9} + \ln {M^2 \over \mu^2}+{11 \over 12}\ln {k^2 \over M^2} + O(1/k^2) \Big] .
\end{align}
For more details of the calculations, see \cite{Kondo00}.

%\newpage
%%%%%%%%%%%%%%%%%%%%%%%%%%%%%%%%%%%%%%%%%%%%%%%%%%
%%%%%Acknowledgments
%%%%%%%%%%%%%%%%%%%%%%%%%%%%%%%%%%%%%%%%%%%%%%%%%%
\section*{Acknowledgments}
The author would like to thank Nora Brambilla and John Field for sending valuable comments, Hideo Suganuma for giving valuable comments on this paper and Toru Shinohara for pointing out careless errors in the original version. 
This work is supported by Sumitomo Foundations, 
Grant-in-Aid for Scientific Research (C)14540243 from Japan Society for the Promotion of Science (JSPS), 
and in part by Grant-in-Aid for Scientific Research on Priority Areas (B)13135203 from
The Ministry of Education, Culture, Sports, Science and Technology (MEXT).

%\newpage

\end{document}